\documentstyle[12pt,aaspp4]{article}
\makeatletter
\let\DOTSI\relax
\def\RIfM@{\relax\ifmmode}%
\def\FN@{\futurelet\next}%
\newcount\intno@
\def\iint{\DOTSI\intno@\tw@\FN@\ints@}%
\def\iiint{\DOTSI\intno@\thr@@\FN@\ints@}%
\def\iiiint{\DOTSI\intno@4 \FN@\ints@}%
\def\idotsint{\DOTSI\intno@\z@\FN@\ints@}%
\def\ints@{\findlimits@\ints@@}%
\newif\iflimtoken@
\newif\iflimits@
\def\findlimits@{\limtoken@true\ifx\next\limits\limits@true
 \else\ifx\next\nolimits\limits@false\else
 \limtoken@false\ifx\ilimits@\nolimits\limits@false\else
 \ifinner\limits@false\else\limits@true\fi\fi\fi\fi}%
\def\multint@{\int\ifnum\intno@=\z@\intdots@                                %1
 \else\intkern@\fi                                                          %2
 \ifnum\intno@>\tw@\int\intkern@\fi                                         %3
 \ifnum\intno@>\thr@@\int\intkern@\fi                                       %4
 \int}%                                                                     %5
\def\multintlimits@{\intop\ifnum\intno@=\z@\intdots@\else\intkern@\fi
 \ifnum\intno@>\tw@\intop\intkern@\fi
 \ifnum\intno@>\thr@@\intop\intkern@\fi\intop}%
\def\intic@{\mathchoice{\hskip.5em}{\hskip.4em}{\hskip.4em}{\hskip.4em}}%
\def\negintic@{\mathchoice
 {\hskip-.5em}{\hskip-.4em}{\hskip-.4em}{\hskip-.4em}}%
\def\ints@@{\iflimtoken@                                                    %1
 \def\ints@@@{\iflimits@\negintic@\mathop{\intic@\multintlimits@}\limits    %2
  \else\multint@\nolimits\fi                                                %3
  \eat@}%                                                                   %4
 \else                                                                      %5
 \def\ints@@@{\iflimits@\negintic@
  \mathop{\intic@\multintlimits@}\limits\else
  \multint@\nolimits\fi}\fi\ints@@@}%
\def\intkern@{\mathchoice{\!\!\!}{\!\!}{\!\!}{\!\!}}%
\def\plaincdots@{\mathinner{\cdotp\cdotp\cdotp}}%
\def\intdots@{\mathchoice{\plaincdots@}%
 {{\cdotp}\mkern1.5mu{\cdotp}\mkern1.5mu{\cdotp}}%
 {{\cdotp}\mkern1mu{\cdotp}\mkern1mu{\cdotp}}%
 {{\cdotp}\mkern1mu{\cdotp}\mkern1mu{\cdotp}}}%
\def\rmfam{\z@}%
\newif\iffirstchoice@
\firstchoice@true
\def\textfonti{\the\textfont\@ne}%
\def\textfontii{\the\textfont\tw@}%
\def\text{\RIfM@\expandafter\text@\else\expandafter\text@@\fi}%
\def\text@@#1{\leavevmode\hbox{#1}}%
\def\text@#1{\mathchoice
 {\hbox{\everymath{\displaystyle}\def\textfonti{\the\textfont\@ne}%
  \def\textfontii{\the\textfont\tw@}\textdef@@ T#1}}%
 {\hbox{\firstchoice@false
  \everymath{\textstyle}\def\textfonti{\the\textfont\@ne}%
  \def\textfontii{\the\textfont\tw@}\textdef@@ T#1}}%
 {\hbox{\firstchoice@false
  \everymath{\scriptstyle}\def\textfonti{\the\scriptfont\@ne}%
  \def\textfontii{\the\scriptfont\tw@}\textdef@@ S\rm#1}}%
 {\hbox{\firstchoice@false
  \everymath{\scriptscriptstyle}\def\textfonti
  {\the\scriptscriptfont\@ne}%
  \def\textfontii{\the\scriptscriptfont\tw@}\textdef@@ s\rm#1}}}%
\def\textdef@@#1{\textdef@#1\rm\textdef@#1\bf\textdef@#1\sl\textdef@#1\it}%
\def\DN@{\def\next@}%
\def\eat@#1{}%
\def\textdef@#1#2{%
 \DN@{\csname\expandafter\eat@\string#2fam\endcsname}%
 \if S#1\edef#2{\the\scriptfont\next@\relax}%
 \else\if s#1\edef#2{\the\scriptscriptfont\next@\relax}%
 \else\edef#2{\the\textfont\next@\relax}\fi\fi}%
\def\Let@{\relax\iffalse{\fi\let\\=\cr\iffalse}\fi}%
\def\vspace@{\def\vspace##1{\crcr\noalign{\vskip##1\relax}}}%
\def\multilimits@{\bgroup\vspace@\Let@
 \baselineskip\fontdimen10 \scriptfont\tw@
 \advance\baselineskip\fontdimen12 \scriptfont\tw@
 \lineskip\thr@@\fontdimen8 \scriptfont\thr@@
 \lineskiplimit\lineskip
 \vbox\bgroup\ialign\bgroup\hfil$\m@th\scriptstyle{##}$\hfil\crcr}%
\def\Sb{_\multilimits@}%
\def\endSb{\crcr\egroup\egroup\egroup}%
\def\Sp{^\multilimits@}%

\newdimen\ex@
\def\rightarrowfill@#1{$#1\m@th\mathord-\mkern-6mu\cleaders
 \hbox{$#1\mkern-2mu\mathord-\mkern-2mu$}\hfill
 \mkern-6mu\mathord\rightarrow$}%
\def\leftarrowfill@#1{$#1\m@th\mathord\leftarrow\mkern-6mu\cleaders
 \hbox{$#1\mkern-2mu\mathord-\mkern-2mu$}\hfill\mkern-6mu\mathord-$}%
\def\leftrightarrowfill@#1{$#1\m@th\mathord\leftarrow\mkern-6mu\cleaders
 \hbox{$#1\mkern-2mu\mathord-\mkern-2mu$}\hfill
 \mkern-6mu\mathord\rightarrow$}%
\def\overrightarrow{\mathpalette\overrightarrow@}%
\def\overrightarrow@#1#2{\vbox{\ialign{##\crcr\rightarrowfill@#1\crcr
 \noalign{\kern-\ex@\nointerlineskip}$\m@th\hfil#1#2\hfil$\crcr}}}%

\def\overleftarrow{\mathpalette\overleftarrow@}%
\def\overleftarrow@#1#2{\vbox{\ialign{##\crcr\leftarrowfill@#1\crcr
 \noalign{\kern-\ex@\nointerlineskip}$\m@th\hfil#1#2\hfil$\crcr}}}%
\def\overleftrightarrow{\mathpalette\overleftrightarrow@}%
\def\overleftrightarrow@#1#2{\vbox{\ialign{##\crcr\leftrightarrowfill@#1\crcr
 \noalign{\kern-\ex@\nointerlineskip}$\m@th\hfil#1#2\hfil$\crcr}}}%
\def\underrightarrow{\mathpalette\underrightarrow@}%
\def\underrightarrow@#1#2{\vtop{\ialign{##\crcr$\m@th\hfil#1#2\hfil$\crcr
 \noalign{\nointerlineskip}\rightarrowfill@#1\crcr}}}%

\def\underleftarrow{\mathpalette\underleftarrow@}%
\def\underleftarrow@#1#2{\vtop{\ialign{##\crcr$\m@th\hfil#1#2\hfil$\crcr
 \noalign{\nointerlineskip}\leftarrowfill@#1\crcr}}}%
\def\underleftrightarrow{\mathpalette\underleftrightarrow@}%
\def\underleftrightarrow@#1#2{\vtop{\ialign{##\crcr$\m@th\hfil#1#2\hfil$\crcr
 \noalign{\nointerlineskip}\leftrightarrowfill@#1\crcr}}}%
\newcount\GRAPHICSTYPE
\GRAPHICSTYPE=\z@
\def\GRAPHICSPS#1{%
 \ifcase\GRAPHICSTYPE%\GRAPHICSTYPE=0
  ps: #1%
 \or%\GRAPHICSTYPE=1
  language "PS", include "#1"%
 \or%\GRAPHICSTYPE=2
  #1%
 \fi
}%
\def\graffile#1#2#3#4{%
 \ifnum\GRAPHICSTYPE=\tw@
  \@ifundefined{psfig}{\input psfig.tex}{}%
  \psfig{file=#1, height=#3, width=#2}%
 \else
  \leavevmode\raise -#4 \hbox{%
   \raise #3 \hbox{\rule{0.003in}{0.003in}\special{#1}}%
   }%
  {\raise -#4 \hbox to #2 {\vrule height#3 width\z@ depth\z@\hfil}}%
 \fi
}%
\def\draftbox#1#2#3#4{%
 \leavevmode\raise -#4 \hbox{%
  \frame{\rlap{\protect\tiny #1}\hbox to #2%
   {\vrule height#3 width\z@ depth\z@\hfil}%
  }%
 }%
}%
\newcount\draft
\draft=\z@
\def\GRAPHIC#1#2#3#4#5{%
 \ifnum\draft=\@ne\draftbox{#2}{#3}{#4}{#5}%
  \else\graffile{#1}{#3}{#4}{#5}%
  \fi
 }%
\def\addtoLaTeXparams#1{\edef\LaTeXparams{\LaTeXparams #1}}%
\def\doFRAMEparams#1{\readFRAMEparams#1\end}%
\def\readFRAMEparams#1{%
 \ifx#1\end%
  \let\next=\relax
  \else
  \ifx#1i\dispkind=\z@\fi
  \ifx#1d\dispkind=\@ne\fi
  \ifx#1f\dispkind=\tw@\fi
  \ifx#1t\addtoLaTeXparams{t}\fi
  \ifx#1b\addtoLaTeXparams{b}\fi
  \ifx#1p\addtoLaTeXparams{p}\fi
  \ifx#1h\addtoLaTeXparams{h}\fi
  \let\next=\readFRAMEparams
  \fi
 \next
 }%
\def\IFRAME#1#2#3#4#5{\GRAPHIC{#5}{#4}{#1}{#2}{#3}}%
\def\DFRAME#1#2#3#4{%
 \begin{center}\GRAPHIC{#4}{#3}{#1}{#2}{\z@}\end{center}%
 }%
\def\FFRAME#1#2#3#4#5#6#7{%
 \begin{figure}[#1]%
  \begin{center}\GRAPHIC{#7}{#6}{#2}{#3}{\z@}\end{center}%
  \caption{\label{#5}#4}%
  \end{figure}%
 }%
\newcount\dispkind%
\def\FRAME#1#2#3#4#5#6#7#8{%
 \def\LaTeXparams{}%
 \dispkind=\z@
 \def\LaTeXparams{}%
 \doFRAMEparams{#1}%
 \ifnum\dispkind=\z@\IFRAME{#2}{#3}{#4}{#7}{#8}\else
  \ifnum\dispkind=\@ne\DFRAME{#2}{#3}{#7}{#8}\else
   \ifnum\dispkind=\tw@
    \edef\@tempa{\noexpand\FFRAME{\LaTeXparams}}%
    \@tempa{#2}{#3}{#5}{#6}{#7}{#8}%
    \fi
   \fi
  \fi
 }%
\long\def\QQQ#1#2{\long\expandafter\def\csname#1\endcsname{#2}}%
\def\QTP#1{}%
\long\def\QQA#1#2{}%
\def\QTR#1#2{{\csname#1\endcsname #2}}%(gp) Is this the best?
\def\EXPAND#1[#2]#3{}%
\def\NOEXPAND#1[#2]#3{}%
\def\LaTeXparent#1{}%
\def\QTagDef#1#2#3{}%
\def\QQfnmark#1{\footnotemark}

\@ifundefined{INDEX}{\def\INDEX#1#2{}{}}{}%
\@ifundefined{SUBINDEX}{\def\SUBINDEX#1#2#3{}{}{}}{}%
\def\initial#1{\bigbreak{\raggedright\large\bf #1}\kern 2\p@\penalty3000}%
\@ifundefined{abstract}{%
 \def\abstract{%
  \if@twocolumn
   \section*{Abstract (Not appropriate in this style!)}%
   \else \small 
   \begin{center}{\bf Abstract\vspace{-.5em}\vspace{\z@}}\end{center}%
   \quotation 
   \fi
  }%
 }{%
 }%
\@ifundefined{endabstract}{\def\endabstract
  {\if@twocolumn\else\endquotation\fi}}{}%
\@ifundefined{maketitle}{\def\maketitle#1{}}{}%
\@ifundefined{affiliation}{\def\affiliation#1{}}{}%
\@ifundefined{proof}{}{}%
\@ifundefined{endproof}{}{}%
\@ifundefined{newfield}{\def\newfield#1#2{}}{}%
\@ifundefined{chapter}{\def\chapter#1{\par(Chapter head:)#1\par }%
 \newcount\c@chapter}{}%
\@ifundefined{part}{\def\part#1{\par(Part head:)#1\par }}{}%
\@ifundefined{section}{\def\section#1{\par(Section head:)#1\par }}{}%
\@ifundefined{subsection}{\def\subsection#1%
 {\par(Subsection head:)#1\par }}{}%
\@ifundefined{subsubsection}{\def\subsubsection#1%
 {\par(Subsubsection head:)#1\par }}{}%
\@ifundefined{paragraph}{\def\paragraph#1%
 {\par(Subsubsubsection head:)#1\par }}{}%
\@ifundefined{subparagraph}{\def\subparagraph#1%
 {\par(Subsubsubsubsection head:)#1\par }}{}%
\@ifundefined{therefore}{}{}%
\@ifundefined{backepsilon}{}{}%
\@ifundefined{yen}{}{}%
\@ifundefined{registered}{\def\registered{\relax\ifmmode{}\r@gistered
                                                \else$\m@th\r@gistered$\fi}%
 \def\r@gistered{^{\ooalign
  {\hfil\raise.07ex\hbox{$\scriptstyle\rm\text{R}$}\hfil\crcr
  \mathhexbox20D}}}}{}%
\@ifundefined{Eth}{}{}%
\@ifundefined{eth}{}{}%
\@ifundefined{Thorn}{}{}%
\@ifundefined{thorn}{}{}%
\@ifundefined{degree}{}{}%
\def\BibTeX{{\rm B\kern-.05em{\sc i\kern-.025em b}\kern-.08em
    T\kern-.1667em\lower.7ex\hbox{E}\kern-.125emX}}%
\newdimen\theight
\def\Column{%
 \vadjust{\setbox\z@=\hbox{\scriptsize\quad\quad tcol}%
  \theight=\ht\z@\advance\theight by \dp\z@\advance\theight by \lineskip
  \kern -\theight \vbox to \theight{%
   \rightline{\rlap{\box\z@}}%
   \vss
   }%
  }%
 }%
\def\qed{%
 \ifhmode\unskip\nobreak\fi\ifmmode\ifinner\else\hskip5\p@\fi\fi
 \hbox{\hskip5\p@\vrule width4\p@ height6\p@ depth1.5\p@\hskip\p@}%
 }%
\def\miss{\hbox{\vrule height2\p@ width 2\p@ depth\z@}}%
\def\tcol#1{{\baselineskip=6\p@ \vcenter{#1}} \Column}  %
\makeatother

\begin{document}

\setlength{\baselineskip}{24pt}

\title{Dynamics and Structure of Three-Dimensional Poloidally Magnetized 
       Supermagnetosonic Jets}

\author{Philip E. Hardee}  
\affil{Department of Physics \& Astronomy \\ The University of Alabama \\
Tuscaloosa, AL 35487 \\ hardee@venus.astr.ua.edu}

\author{David A. Clarke}
\affil{Department of Astronomy \& Physics \\ Saint Mary's University \\
 Halifax, Nova Scotia, B3H 3C3 \\ dclarke@ap.stmarys.ca} 

\and

\author{Alexander Rosen}
\affil{Department of Physics \& Astronomy \\ The University of Alabama \\ 
 Tuscaloosa, AL 35487 \\ rosen@eclipse.astr.ua.edu}

\newpage
\setlength{\baselineskip}{12pt}

\begin{abstract}

A set of three-dimensional magnetohydrodynamical simulations of
supermagnetosonic magnetized jets has been performed. The jets contain
an equipartition primarily poloidal magnetic field and the effect of
jet density on jet dynamics and structure is evaluated. The jet is
precessed at the origin to break the symmetry and to excite
Kelvin-Helmholtz unstable helical modes. In the linear limit observed
structures are similar in all simulations and can be produced by
structures predicted to arise as a result of instability.  The
amplitude of various unstable modes is evaluated. Most unstable modes
do not reach the maximum amplitudes estimated from the linear theory by
computing displacement surfaces associated with the modes.
Surprisingly, even these large amplitude distortions are fit reasonably
well by displacement surfaces computed from the linear theory.  Large
amplitude helical and elliptical distortions lead to significant
differences in the nonlinear development of the jets as a function of
the jet density. Jets less dense than the surrounding medium entrain
material, lose energy through shock heating and slow down relatively
rapidly once large amplitude distortions develop as a result of
instability. Jets more dense than the surrounding medium lose much less
energy as they entrain and accelerate the surrounding medium. The dense
jet maintains a high speed spine which exhibits large amplitude helical
twisting and elliptical distortion over considerable distance without
disruption of internal jet structures as happens for the less dense
jets. This dense high speed spine is surrounded by a less dense sheath
consisting of slower moving jet fluid and magnetic field mixed with the
external medium.  Simulated synchrotron intensity and fractional
polarization images from these calculations provide a considerably
improved connection between simulation results and jet observations
than do images made using the fluid variables alone.  Intensity
structure in the dense jet simulation appears remarkably similar to
structure observed in the Cygnus A jet.  These simulations suggest that
the extended jets in high power radio sources propagate to such large
distances without disruption by entrainment because they are surrounded
by a lobe or cocoon whose density is less than the jet density.

\end{abstract}


\newpage
\section{Introduction}

The highly collimated outflows from extragalactic radio sources, and
protostellar systems, e.g., HH\,34 (Burke, Mundt, \& Ray 1988), and
probably also the galactic superluminals GRS\,1915+105 (Mirabel \&
Rodr\'\i guez 1994) and GRO\,J1655--40 (Tingay et al. 1995; Hjellming \&
Rupen 1995) are subject to instability that can produce organized
helical structures and will lead to momentum and mass exchange with the
external environment. Observation of extragalactic jets shows that jets
can propagate many tens of times their diameter while remaining highly
collimated, e.g., Cygnus A (Dreher, Carilli, \& Perley 1987) and NGC\,6251 (Perley, Bridle, \& Willis 1984), or lose their narrow collimation
after a relatively short distance, e.g., M\,84 (Laing \& Bridle 1987).
Observation also shows that jets can show bulk large scale sinusoidal
distortion, e.g., 3C\,449 (Perley \& Cornwell 1984) or exhibit smaller
scale apparent sinusoidal oscillations, e.g., HH\,34 and GRO\,J1655--40,
which, if the jets are not ballistic, may be related to flow
instability possibly triggered by precession or orbital
motion associated with the central engine. In this case it is hoped
that observed jet structures can be related to jet velocity, density,
and magnetic fields via numerical simulations, stability theory, and
modeling. This is of particular importance to extragalactic jets or
those galactic jets where the continuum nature of synchrotron and
inverse Compton emission provides no direct measure of these quantities
and whose apparent motions may reflect a wave pattern or shock speed
different from the speed of the underlying flow.

Theoretical results and laboratory experiments (see Bicknell 1984,
1986a, b) and three-dimensional numerical simulations (Hardee, Clarke,
\& Howell 1995) have indicated that momentum and mass exchange with an
external environment more dense than a fluid jet results in a
relatively rapid loss of the initial high collimation, and the outward
flow broadens and slows as denser external material is mixed with and
accelerated by the lighter jet fluid. On the other hand, extragalactic
jets in FR~II type radio sources that remain highly collimated are
those surrounded by a lobe or cocoon almost certainly less dense than
the jet (Norman et al. 1984). In addition, the galactic protostellar
jets are perhaps ten times more dense than the immediately surrounding
medium. It is also possible that the inclusion of magnetic fields
reduces momentum and mass exchange with the external environment. That
a sufficiently supermagnetosonic magnetized jet denser than the
immediately surrounding medium might mitigate the Kelvin-Helmholtz
disruption of extragalactic jets was suggested by a three-dimensional
numerical simulation performed by Hardee \& Clarke (1995). In this
paper we report on results obtained from three numerical simulations of
magnetized cylindrical jets which are initialized across a Cartesian
grid. Thus, these simulations are relevant to astrophysical jets far
behind the propagating jet working surface. The principal difference
between the simulations is the different jet density. The inclusion of
magnetic fields in these simulations allows an examination of the
dynamical effects that magnetic fields have on stability through
comparison with previous purely fluid jet simulations. The magnetic
fields also allow computation of total intensity and fractional
polarization images that provide a better connection to the
observations than do the fluid variables.

\newpage
\section{Numerical Simulations}

Simulations were performed using the three-dimensional MHD code
ZEUS-3D, an Eulerian finite-difference code using the recently
developed Consistent Method of Characteristics (CMoC). The strength of
the CMoC lies in its ability to solve the transverse momentum transport
and magnetic induction equations simultaneously and in a {\it planar
split} fashion. Additional details may be found in Clarke (1996). The
development of CMoC has allowed us to perform 3-D MHD calculations such
as those presented here with some confidence. Interpolations are
carried out by a second-order accurate monotonic upwinded time-centered
scheme (van Leer 1977) and a von-Neumann Richtmyer artificial viscosity
is used to stabilize shocks. The MHD test suites as described by Stone
et al. (1992) and Clarke (1996) were used to establish the reliability
of the techniques.

\subsection{Initialization}

Simulations are initialized by establishing a cylindrical jet
across a three-dimensional Cartesian grid resolved into 130 $\times $
130 $\times $ 325 zones. Thirty uniform zones span the jet diameter
along the transverse Cartesian axes ($y$-axis and $z$-axis). Outside
the jet the grid zones are ratioed where each subsequent zone increases
in size by a factor 1.05.  Altogether the 130 zones along the
transverse Cartesian axes span a total distance of 30 jet radii. Along
the $x$-axis 225 uniform zones span a distance of 30 jet radii outwards
from the jet origin. An additional 100 ratioed zones span an additional
distance of 30 jet radii where each subsequent zone increases in size
by a factor 1.0148.  Outflow boundary conditions are used at all
boundaries except where the jet enters the grid where inflow boundary
conditions are used. The use of a non-uniform grid such as we are
employing here has been shown to have the beneficial effect of reducing
reflections off the grid boundaries as a result of increased
dissipation of disturbances by the larger grid zones near the grid
boundaries (Bodo et al. 1995).  With 15 zones across the jet radius in
the transverse direction and 7.5 zones per jet radius in the axial
direction we should be able to detect structures with length scales larger
than about 25\% of the jet radius in the transverse direction and 50\% of
the jet radius in the axial direction.

The ``equilibrium jets'' (as opposed to ``propagating jets'') shown in
\S~2.2--2.4 are initialized with a uniform density $\rho _{jt}$ across
the computational grid within an external medium of uniform density
$\rho _{ex}$. The thermal pressure in the jet is initialized at
$p_{jt}^{th}=0.54p_{ex}$ where $p_{ex}$ is the pressure in the uniform
external medium. The magnetic field in the jet is initialized with a
uniform poloidal component, B$_x$, corresponding to a magnetic pressure
$p_{jt}^{B_x}=0.469p_{ex}$. In the absence of a toroidal magnetic
component this leads to a small ($\sim 1\%$) overpressure of the jet
compared to the external medium. However, a toroidal magnetic component
is included in the jet and its variation with radius is shown in Figure
1.  The toroidal component has a maximum value of $B_\phi ^{pk}\simeq
0.339B_x$, gives a helical twist to the jet magnetic field and provides
a confining magnetic pressure which at maximum value is equal to $\sim
11.6\%$ of the poloidal magnetic pressure.  The total transverse static
jet pressure varies from greater than the external pressure by about
1\% at jet center and near to the jet surface to about 5\% less than
the external pressure where the toroidal magnetic component is a
maximum. The helical pitch of the magnetic field is also indicated in
Figure 1. The maximum pitch of $\sim 18.7^{\arcdeg}$ at a radial
distance of 0.8R corresponds to a 360$^{\arcdeg}$ twist around the jet
in an axial distance of $\sim 14.8R$.  This primarily poloidal magnetic
field configuration of the three simulations shown in \S~2.2--2.4 has
been chosen so that structures arising in the numerical simulations can
be directly compared to structures predicted by linear stability theory
(\S~3 and \S~4).  Additional simulations with different magnetic field
configurations and strengths have also been performed but a detailed 
analysis of the results of these additional simulations is beyond the 
scope of this paper.

The initial conditions for the three simulations are given in Table 1.
In all simulations the external density and temperature are the same.
The jet density is varied, and the jet temperature is varied
appropriately to maintain a constant jet thermal pressure. The internal
sonic, Alfv\'enic and magnetosonic Mach numbers are defined as
$M_{jt}\equiv u/a_{jt}$, $M_{jt}^A\equiv u/V_{A,jt}$ and
$M_{jt}^{ms}\equiv u/(a_{jt}^2+V_{A,jt}^2)^{1/2}$, respectively, where
$a_{jt}$, and $V_{A,jt}$ are the jet sound speed, and the poloidal
Alfv\'en speed, respectively, and we define a jet magnetosonic speed as
$a_{jt}^{ms}\equiv (a_{jt}^2+V_{A,jt}^2)^{1/2}$. The jet speed is
initialized to keep the jet magnetosonic Mach number within a
relatively small range.  For sufficiently supermagnetosonic jets such
as we study here, the theoretically predicted stability behavior is
primarily a function of the jet magnetosonic Mach number.  Differences
between the simulations are expected to be the result of nonlinear
behavior primarily related to the density difference and secondarily
related to the velocity difference.  Some difference between the
present and previous purely fluid simulations (Hardee, Clarke, \&
Howell 1995) may be related to the addition of the relatively strong
equipartition magnetic field.  With jet speeds corresponding to
magnetosonic Mach numbers, $M_{jt}^{ms}$, of (a) $\sim 5.95$, (b) $\sim
4.90$ and (c) $\sim 7.70$, one flow through time (i.e., the time
required for jet material to flow completely across the computational
grid) is $\tau _f\equiv t_fa_{ex}/R\simeq $ (a) 4.8, (b) 11.6 and (c)
14.8, where $a_{ex}$ is the sound speed in the external medium.

In the simulations presented here the cylindrical jet is driven by a
periodic precession of the jet velocity, $u$, at the inlet with an
angle of 0.005 radian relative to the $x$-axis in a counterclockwise
sense when viewed outwards from the inlet and with an angular frequency
(a \& b) $\omega = 1.6a_{ex}/R$, and (c) $\omega = 1.2a_{ex}/R$. These
precessional frequencies are near the theoretically predicted maximally
unstable frequency associated with helical twisting of the jet, and
serve to excite other Kelvin-Helmholtz unstable helical modes. With
these precession frequencies one complete precessional period occurs in
a time (a \& b) $\tau_p \equiv t_pa_{ex}/R \simeq 3.93$, and (c)
$\tau_p \simeq 5.24$.  The initial transverse motion imparted to the
jet by the precession in the simulations is (a) $ \simeq
0.030a_{jt}^{ms}$ or $\simeq 0.062a_{ex}$, (b) $\simeq
0.025a_{jt}^{ms}$ or $\simeq 0.026a_{ex}$ and (c) $\simeq
0.039a_{jt}^{ms}$ or $\simeq 0.020a_{ex}$, and is well within the
linear regime of a small perturbation. The direction of precession
induces a helical twist in the same sense as that of the magnetic field
helicity so that the toroidal component of the field has relatively
little influence on the development of unstable growing waves, i.e.,
helical wavefronts are at shallow angles to the helically twisted
magnetic field (Hardee et al.\ 1992), and this choice
also guarantees that the magnetic field experiences the maximum
compressive effects associated with the helical perturbations induced
by the precession. After dynamical times $\tau _d \equiv t_da_{ex}/R=$
(a) 17, (b) 28, and (c) 36 the numerical simulations have reached a
quasi-steady state out to axial distances in excess of 50R.  These
dynamical times correspond to more than 4 precession periods and about
3 flow through times through an axial distance of 50R.

Images from the simulations consist of magnetic pressure cross
sections, and line-of-sight integrations corresponding to total
synchrotron intensity and fractional polarization. The cross
section images of dimension $10R \times 10R$ are oriented with the $y$-axis
and $z$-axis in the vertical and horizontal directions, respectively,
and with the $x$-axis into the page, i.e., we look outwards along the
flow direction.  To generate intensity and fractional polarization
images, an emissivity is defined by $p^{th}$($B\sin \theta $)$^{3/2}$
where $\theta $ is the angle made by the local magnetic field with
respect to the line-of-sight.  This emissivity can be shown to mimic
synchrotron emission from a system in which the energy and number
densities of the relativistic particles are proportional to the energy
and number densities of the thermal fluid. This simplistic assumption
is necessary when the relativistic particles are not explicitly tracked
(Clarke, Norman, \& Burns 1989). The Stokes parameters are computed
from line-of-sight integrations along the $z$-axis, and from these
parameters the simulated intensity and fractional polarization images
are generated. The displayed images have a dimension of $20R\times
60R$. Recall that the computational grid has dimension $30R\times
30R\times 60R$. We note that an image formed by a line-of-sight
integration of the magnetic pressure is very similar in appearance to
the total intensity image. In addition, the magnetic pressure cross
sections and line-of-sight images reveal the extent of
jet spreading and structure as only the jet fluid is magnetized.

\subsection{Magnetic Pressure Cross Sections and Line-of-Sight Images}

Figures 2, 3, 4 and 5 each contain eight magnetic pressure cross
section images, at axial distances from $6R$ to $48R$ in increments of
$6R$, and two line-of-sight images.  Figures 2 and 3 are from
simulations A (light jet) and B (equal density jet), respectively.
Figures 4 and 5 are from simulation C (heavy jet) at dynamical times
$\tau_d$ = 34, and $\tau_d$ = 36, respectively.  In simulation A
(Figure 2), the images show that out to axial distances of $\sim 24R$
the jet remains well collimated and shows evidence for organized
internal jet structure.  Fluting of the jet surface is apparent in the
cross section at 12$R$ and the cross sections show some jet flattening
with a twist through $\sim 180^{\arcdeg }$ between 12$R$ and 24$R$. The
line-of-sight intensity image shows twisted bright filaments that are
related to fluting and flattening of the jet.  In simulation B (Figure
3), the jet remains well collimated and shows evidence for organized
internal jet structure out to axial distances of $\sim 29R$. Surface
fluting is very pronounced in the cross section at 12$R$ (seven ripples
are apparent) and the cross sections show evidence for flattening
particularly at an axial distance of 30$R$. Cross sections at other
dynamical times suggest a flattening which twists through 180$^{\arcdeg
}$ over a length scale of $\sim 13.5R$. At axial distances between 16
-- 25$R$ dual filamentation twists through $180^{\arcdeg }$ over a
length scale of $\sim 4.5R$ (see Figure 8 in Hardee 1995).

In simulation C (Figures 4 and 5), the images suggest that the flow
remains relatively highly collimated across the entire computational
grid. For details of the structures seen at $\tau _d = 34$ refer to
Hardee \& Clarke (1995). Small scale surface fluting like that seen in
the less dense jets is not apparent in these images but instead the jet
appears square in cross sections out to $\sim 24R$.  Animation of the
cross sections reveals that the square distortion is only partially an
effect of the Cartesian grid and that the distortion rotates temporally
and spatially.   Interestingly, the square distortion does not show a
smooth rotation in temporal animations, but shows a square pattern that
changes temporally in orientation in 45$^{\arcdeg }$ increments after
passing through a fluted cross section stage. The lack of change in
orientation between the panels from 6$R$ to 24$R$ and between the two
figures is the serendipitous result of a 90$^{\arcdeg}$ spatial
rotation with wavelength $\approx 6R$ (the spatial separation between
the panels) and a 90$^{\arcdeg}$ temporal rotation which occurs in
$\Delta \tau _d\approx 2$ (the temporal separation between the
figures). In \S 4 we will see that an additional triangular distortion
(suggested by the cross section at 18R in Figure 4) with a
120$^{\arcdeg}$ spatial rotation in a wavelength $\approx 9R$ is also
required at these axial distances.

A small amplitude sinusoidal oscillation in the line-of-sight images
(helical twist of the jet) is evident in simulations A and B at axial
distances $< 30R$ with wavelengths $\approx 20R$, and $\approx 11.5R$,
respectively.  The helical twist moves outwards with a wave speed
$v_w \approx 0.30u$ and $v_w \approx 0.54u$ in simulations A and B,
respectively.  At larger axial distances in simulation A, there is
evidence for a helical twist but with wavelength decreasing to $\sim
16R$.  In simulation B, at times earlier than shown in the present
image we find evidence for sinusoidal oscillation with a wavelength
$\approx 5.5R$ at axial distances $15R-30R$ that are not readily
identifiable with features in the cross sections.  Simulation C
exhibits a relatively well defined large scale sinusoidal oscillation
with a wavelength $\approx 14R$ across most of the computational grid.
The oscillation moves outwards by $\sim 6R$ between the two images.
This motion corresponds to a wave speed of $v_w \approx 0.74u$.

All jets become less well-collimated in the outer half of the
computational grid. In simulation A, beyond about 27$R$ organized
structure is disrupted and jet material can be found up to a distance
of $\sim 5R$ from the jet axis at large axial distances. For example,
magnetized jet material fills much of the $10R \times 10R$ cross
section at an axial distance of 48$R$ and is clearly mixed with
unmagnetized external material.  In simulation B, the flow begins to
broaden significantly at an axial distance of about 30$R$ and beyond this
distance organized structure is much less well defined.  While
magnetized jet material can be found up to a distance of $\sim 5R$ from
the jet axis and magnetized jet material fills a considerable portion
of the $10R \times 10R$ cross section at an axial distance of
48$R$, there is more central concentration than in simulation A. There is
a suggestion of dual magnetic filaments in the cross section at 48$R$ in
simulation B.  In simulation C, magnetized jet material can be found
only up to a distance of $\sim 2.5R$ in the cross section images and
this reflects the strong central concentration of the jet although in
this case not the true extent of mixing.

In simulation C,  organized internal structure and collimation beyond
an axial distance of 30$R$ remain even after a helical twist and jet
flattening have grown to significant amplitudes.  Jet flattening
appears in the line-of-sight images as alternating broad and narrow
regions which move outwards with a wave speed similar to the wave speed
of the large scale sinusoidal oscillation.  The cross section images
suggest that magnetic filaments lie at the ends of the long axis of
the flattened jet.  At a fixed distance the flattened jet rotates
counterclockwise by almost 90$^{\arcdeg}$ between Figures 4 and 5 and
rotates in the same sense as jet precession at the origin.  We conclude
that the jet is subject to flattening and dual filamentation that
twists through 180$^{\arcdeg}$ with a wavelength $\lambda /R \approx
14$. The flattening goes through a nearly 360$^{\arcdeg}$ rotation
without disruption.   Note that the 360$^{\arcdeg}$ helical twist
induced by precession at the jet origin occurs over approximately the
same axial distance as the 180$^{\arcdeg}$ twist in jet flattening.
There is also some evidence of filament twisting with wavelengths
$\approx 4R$ inside the jet. Additional complex structures on
relatively short scale lengths are also revealed in the intensity
images at distances between $12R-24R$ and appear to be associated with
structures helically twisted around the jet both at the jet surface and
in the jet interior but are not easily related to features in the cross
sections.

In simulation C, the bright diagonal structures crossing the jet from
left-bottom to right-top are on the back side of the jet if they lie
parallel to the large scale helical twist of the jet. The locations of
bright diagonal filaments in the total intensity images appear to
correspond to those locations where magnetic filaments associated with
jet flattening are aligned with the leading edge of jet helicity
induced by precession at the origin.   Note the line-of-sight merging
of the twin diagonal features in the intensity image in Figure 4 at
axial distances of $\sim 24R$ and $\sim 38R$ into single diagonal
features at axial distances of $\sim 32R$ and $\sim 44R$ in the
intensity image in Figure 5. This change implies that different
structures, e.g., helical twisting and jet flattening, do not move with
exactly the same speed.  Note that other small scale cross section and
intensity features change significantly without simple relation between
Figures 4 and 5.

In all simulations out to $\sim 15R$ the fractional polarization is
higher at the jet edge indicative of the high ordering of the poloidal
magnetic field at the jet edges, and is lower in the interior
indicative of the helical twisting of the field.  In
general, the fractional polarization declines rapidly beyond an axial
distance of 27$R$ in simulation A, and declines beyond 30$R$ for the
denser jets in simulations B and C.  However, a central spine is
evident in simulation B out to about 42$R$ in both the total intensity
and fractional polarization images and in simulation C out to about
54$R$. Low fractional polarization at large axial distances occurs as
the line-of-sight passes through many zones with different magnetic
field orientation although the magnetic field may be ordered on size
scales larger than one zone.  The diagonal structures seen in the
intensity image in simulation C are relatively highly polarized
suggesting strong ordered magnetic fields in these filaments. The
increase of fractional polarization at the outer edges of the
magnetized sheath is caused by fewer zones contributing to the
line-of-sight integration and may not indicate a higher level of
organized structure in the magnetic field.  However, the filamentary
features and ripples at the outer edges of the sheath, which are seen
in both the fractional polarization and intensity images of all three
simulations, suggest some overall organization in fluid motions at the
outer edges of the magnetized sheath.

At an axial distance of 48$R$, the fractional polarization images show
that magnetized jet material spans a distance of $\sim 10R$, $\sim
8.5R$ and $\sim 8R$, and the intensity images indicate a distance of
$\sim 8.5R$, $\sim 7R$, and $\sim 5R$ in simulations A, B and C,
respectively. The difference in the sizes indicates a reduction in
mixing and an increasing central concentration as the jet density is
increased and the jet velocity is decreased.  In fact, in simulation C,
magnetized jet material appears to fill only a small portion of the
cross section at 48$R$ because the amount of magnetized material and
the associated magnetic field and magnetic pressure are so low far from
the central concentration.

Organized structures found in the line-of-sight and cross section
images are summarized below.

1. In simulation A, images show evidence for pinching ($\lambda _p/R \sim
6$), helical twisting ($\lambda _h/R\sim 20$) and elliptical distortion
($\lambda _e/R \sim 12$).

2. In simulation B, images show evidence for helical twisting ($\lambda
_h/R \sim 12$), internal transverse oscillation ($\lambda _h/R \sim
5$), elliptical distortion ($\lambda _e/R \sim 13.5$), and internal
dual twisted filaments that we identify with elliptical distortion
($\lambda _e/R \sim 4.5$).

3. In simulation C, images show evidence for helical twisting ($\lambda
_h/R \sim 14$), elliptical distortion ($\lambda _e/R \sim 14$),
internal dual twisted filaments ($\lambda _e/R \sim 4$), triangular
distortion ($\lambda _t/R \sim 9$), and square distortion ($\lambda
_s/R \sim 6$).

4. All structures move with a comparable but not identical speed with
$v_w \approx 0.30u$, $v_w \approx 0.54u$ and $v_w \approx 0.74u$ in
simulations A, B and C, respectively.

\subsection{Structure Along the Jet Axis}

Axial profiles of density, temperature, sonic Mach number ($M_{jt}$),
and the three velocity components are shown for simulations A, B and C
in Figures 6, 7 and 8, respectively.  The axial velocity remains
constant out to axial distances of $\sim 27R$, $\sim 29R$, and $\sim
32R$ in simulations A, B and C, respectively, after which there are one
or more significant dips.  Temperature and density increases associated
with the dips suggest that these features are shocks.  The dips appear
periodically with wavelengths of $\sim 8R $, $\sim 9R$ and $\sim 14R$,
in simulations A, B and C, respectively. Several bright knots or
filaments in the intensity images can be associated with the dips in
the axial velocity plots.  For example, a bright knot at $\sim 27R$ in
simulation A (see Figure 2), one at $\sim 29R$ in simulation B (see
Figure 3), and the diagonal filaments at $\sim 33R$ and $\sim 45R$ in
simulation C (see Figure 5) are at the locations of dips in the axial
velocity.  The dips are accompanied by a decrease in the average
velocity which falls off more slowly as the jet density is increased.
Typically, the decrease in average jet velocity is accompanied by an
increase in average temperature, and a decrease in the sonic Mach
number.  At large axial distances the light and equal density jets, are
reduced to transonic and mildly supersonic speeds, respectively, while
the dense jet remains highly supersonic. The axial profiles also show
smaller scale axial velocity oscillations, which are most evident in
the in the Mach number plot. These oscillations have wavelengths $\sim
1R$ and $\sim 2R$ at axial distances between $\sim 13-17R$ and $\sim
19-26R$, respectively, in simulation A, and at axial distances $\sim
5-10R$ and $\sim 20-27R$, respectively, in simulation B.  Somewhat
longer wavelengths, $\sim 1.5R$ and $\sim 3.5R$, are found between
$\sim 23-30R$ and $\sim 32-45R$, respectively, in simulation C.

The axial plots also reveal that the transverse velocity contains
oscillatory structure at many different wavelengths:  $\sim 2R$, $\sim
4R$ and $\sim 20R$ in simulation A; $\sim 1.5R$, $\sim 3R$ and $\sim
12R$ in simulation B; $\sim 1.5R$, $\sim 3R$ and $\sim 14R$ in
simulation C.  At the longest wavelength the orthogonal transverse
velocity components are out of phase and are associated with helical
twisting of the jet. In simulation B, there is also evidence for
transverse oscillations at a wavelength of $6-7R$ between axial
distances of $30-50R$.  Typical transverse velocities are $\sim
0.75a_{ex}$ ($\sim 0.4a_{jt}^{ms}$), $\sim 0.4a_{ex}$ ($\sim
0.4a_{jt}^{ms}$) and $\sim 0.2a_{ex}$ ($\sim 0.4a_{jt}^{ms}$) in
simulations A, B and C, respectively.  Maximum transverse velocities
are $\sim 4a_{ex}$ ($\sim 2a_{jt}^{ms}$), $\sim a_{ex}$ ($\sim
a_{jt}^{ms}$) and  $\sim 0.8a_{ex}$ ($\sim 1.5a_{jt}^{ms}$) in
simulations A, B and C, respectively.  In simulation A the maximum
transverse velocity is comparable to the axial velocity.

\subsection{Structure Transverse to the Jet Axis}

Transverse profiles of density, magnetic pressure and axial velocity
along the $y$-axis at axial distances from 12$R$ to 48$R$ in 12$R$
increments are shown in Figures 9, 10 and 11 for simulations A, B and C
(at $\tau_d$ = 36). Note that the vertical scales are not constant.
In simulations A and B, the typical magnetic pressure declines by over
a factor of four and three, respectively, while typical jet densities
decline by no more than about a factor of four and three,
respectively.  As a result the typical Alfv\'en speed decreases
slightly and Alfv\'enic Mach number, $ M_{jt}^A$, increases slightly
with axial distance.  Thus, the light jet is transmagnetosonic and
transonic, and the equal density jet is weakly supermagnetosonic and
weakly supersonic at large axial distances.  In simulation C, the
typical magnetic pressure and density vary only slightly and in a
similar fashion, and the slight drop in axial velocity leads to the
heavy jet remaining supermagnetosonic and supersonic at large axial
distances.

By an axial distance of 12$R$ mixing of jet and external fluid has
begun in all simulations.  The dense jet is still highly collimated
although acceleration of material external to the jet to velocities
$\sim 1.5a_{ex}$ is suggested by the velocity profile.  That there has
been mixing between the fluids is suggested by the small magnetic field
in fluid outside the initial jet radius.  The accelerations (and
associated magnetic fields beyond 1$R$) are smaller in the 12$R$
profiles for the light and equal density jets.  At somewhat larger
axial distances some of the asymmetries in the density and magnetic
pressure profiles are related to the growth of helical twisting.

Farther down the jet at an axial distance of 24$R$ the magnetic
pressure profiles show that magnetized jet material spans a distance of
$\sim 6R$, $\sim 5R$ and $\sim 4R$ in simulations A, B and C,
respectively.  At an axial distance of 48$R$ the magnetic pressure
profile spans a distance of $\sim 10R$, $\sim 7R$ and $\sim 6R$ in
simulations A, B and C, respectively.  In simulation A the spread of
magnetized jet material in the transverse profile corresponds
relatively closely to the spread in the fractional polarization image
(see Figure 2).  In simulation B and C the spread of magnetized jet
material suggested by the transverse profile, corresponds more closely
to the spread in the intensity image than the fractional polarization
image (see Figures 3 and 5) and is evidence for a more centrally
concentrated jet. In simulation C the density, magnetic pressure and
velocity profiles indicate a central spine with diameter comparable to
the initial diameter of the jet.

At an axial distance of 48$R$ the transverse axial velocity profile
shows that the outer portion of the magnetized sheath in simulation A
is nearly at rest and that on average the outwards velocity has dropped
to less than half the velocity at the inlet.  The transverse axial
velocity profile in simulation B shows that the velocity is $\geq 50\%$
of the value at the inlet within a jet diameter, and  $\sim 80\%$ of
the value at the inlet within a spine of width on the order of a jet
radius.  The velocity profile also suggests that most of the magnetized
sheath has a velocity $\approx 0.5a_{ex}$.  In simulation C the
transverse velocity and magnetic profiles suggest a significant
sheath.  The magnetized sheath approximately corresponds to the region
in which flow speeds exceed the external sound speed. At larger
transverse distances outwards flow is $\leq 0.5a_{ex}$ and the lack of
magnetic field suggests that jet fluid has not mixed with external
fluid at these larger transverse distances.

In simulations A and B, considerable heating and expansion have
accompanied mixing of jet and external material.  In simulation A at
and axial distance of 48$R$, the density over the region (diameter of
about 4R) in which the velocity exceeds three times the external sound
speed (roughly half the maximum speed at this distance) is typically
less than half the jet density at the inlet. Outside the magnetized
sheath the medium has a density significantly below the ambient
external density.  In simulation B at 48$R$, the density has decreased
by about a factor of three within the jet spine and the density is
typically decreased by about a factor of two within the magnetized
sheath from the jet density at the inlet. Up to a distance of 2$R$
outside the magnetized sheath, the density remains below 80\% of the
initial jet and ambient density.  In simulation C, there appears to
have been very little heating or expansion from mixing of jet and
ambient fluid.  Within the magnetized sheath the density is comparable
to the initial density in the undisturbed external medium.

All of these simulations, which initially have a strong poloidal magnetic
field, show some evidence for filamentation, and this is most easily
seen in the dense jet simulation.  In Figure 11 at an axial distance
of 36$R$ the magnetic pressure and velocity profiles show some
filamentation into multiple streams outside the central spine.  At
48$R$ the profiles indicate further filamentation into multiple streams
and the magnetic field, largely concentrated in one of the streams, is
very narrow along the $y$-axis.  This narrowing is the reason for the
sharp narrow maximum in the intensity image (see Figure 5) at this
distance. This stream still maintains the initial jet speed.

\subsection{Simulation Summary and Discussion}

In the linear regime close to the inlet all three jets maintain
relatively well organized surface and internal structure on many length
scales both in the images and in the velocity plots.  These structures
include fluting of the jet surface, and helically twisted bright
filaments at the jet surface (possibly associated with the surface
fluting) and in the jet interior and pinching in the jet interior. At
least approximately, smaller scale structures appear nearest to the
inlet with larger scale structures appearing farther from the inlet. In
these simulations, all conducted with comparable magnetosonic Mach
number, the linear development is similar. The smaller scale lesser
amplitude structures disappear as larger scale structures grow to
larger amplitudes farther from the inlet. All features propagate with
comparable but not exactly the same ``wave'' speed.  

The jets in these simulations propagate about half-way across the
computational grid before filamentation and entrainment, in these
simulations associated with jet flattening or helical twisting, become
important.  As the jet density increases, the jet propagates to larger
distances before these nonlinear effects occur. Based on purely
theoretical predictions from a linear analysis (see the following
sections) the light jet (simulation A) with $M_{jt}^{ms}\sim 6$ should
extend about 21\% farther and the heavy jet (simulation C) with $
M_{jt}^{ms}\sim 7.7$ should extend about 57\% farther than the equal
density jet (simulation B) with $M_{jt}^{ms}\sim 4.9$ before nonlinear
effects associated with large scale structures become important. In
fact the light jet does not extend quite as far as the equal density
jet and the heavy jet extends significantly farther than might be
naively expected before the jet shows signs of slowing significantly.
That is, the typical axial velocity remains nearly constant out to
$\sim 27R$, $\sim 29R$, and $\sim 48R$ in simulations A, B, and C,
respectively.

When the light jet loses organized internal structure about half-way
across the computational grid it assumes a plume like appearance with
some residual oscillation associated with helical jet twisting excited
by the precessional motion at the inlet. Transverse motions in the
``plume'' are typically submagnetosonic and no more than one fifth of
the axial motion.  Occasionally, however, transverse motions are
transmagnetosonic and comparable to the outwards motion.  On the other
hand, the heavy jet develops a reasonably pronounced spine-sheath
morphology in which a jet spine maintains the initial jet speed and is
surrounded by a more slowly moving sheath. The jet spine maintains
large scale structure, e.g., jet flattening and helical twisting,
across the entire computational grid. The equal density jet has a
structure that is midway between the light and heavy jets.  We find at
the outer boundary of the computational grid that the maximum flow speed of
the light, equal density, and heavy jets is approximately 50\%, 80\%
and 100\% of the speed at the inlet, respectively.

A control simulation, parameters like simulation C, has been performed
with a precession frequency of $\omega R/a_{ex} = 1$ and a transverse
velocity component of $10^{-6}u$.  This precession provides just enough
perturbation to smooth the jet boundary on the Cartesian grid but
negligible angular momentum.  In this case the jet remains essentially
axisymmetric across the computational grid with complex internal
structure.  Thus, the organized helical structures that we see in the
present simulations do not occur in the absence of a significant
precessional motion at the inlet.

Other numerical simulations have revealed that there is no significant
dynamical difference between a purely poloidal magnetic field
configuration and the present magnetic field configuration.
Additionally, only small differences are seen in simulations with
weaker poloidal magnetic fields.  In general, numerical simulations
indicate that poloidal magnetic field has a relatively small although
not negligible influence on the dynamics and stability of the
collimated flow at a constant supermagnetosonic Mach number and
constant density ratio.  On the other hand, the inclusion of a strong
toroidal magnetic field has been found to have a significant influence
on jet dynamics.  In general, the inclusion of a strong toroidal
magnetic field appears to inhibit the development of the internal
structures and the flattening and fluting surface distortions seen in
the present poloidal magnetic field simulations.  Accompanying the
inhibition of these features is an inhibition in filamentation and a
reduction in entrainment (Rosen et al.\ 1997).  Thus, numerical
simulations indicate that the inclusion of strong toroidal field has a
stabilizing influence (Rosen et al., in preparation).

\section{Jet Stability Theory and Jet Structure}

The stability of a poloidally magnetized cylindrical jet with top hat
profile residing in a uniform unmagnetized medium has been investigated
in several articles (Ray 1981; Ferrari, Trussoni, \& Zaninetti 1981;
Fiedler \& Jones 1984; Bodo et al.\ 1989).  For supermagnetosonic jets
increased poloidal magnetic field strength leads to only a very slight
stabilization relative to a purely fluid jet.  Strong toroidal magnetic
fields, e.g., toroidal magnetic fields of strength comparable to the
present equipartition poloidal magnetic fields, can influence the
stability properties significantly, and lead to increased jet stability
(Appl \& Camenzind 1992).  However, inclusion of a small non-uniform
toroidal magnetic field component such as is present in these numerical
simulations does not significantly modify results obtained from the
following linear stability analysis incorporating only poloidal
magnetic fields.

In cylindrical geometry a random perturbation to the equilibrium jet
can be considered to consist of Fourier components of the form
$f_1(r,\phi,x) = f_1(r)\exp [i(kx \pm n\phi-\omega t)]$ where the flow
is along the $x$-axis.  These components propagate as waves with a
dispersion relation given by equation (A6) in the Appendix.  The normal
modes of a cylindrical jet involve pinch ($n=0$), helical ($n=1$),
elliptical ($n=2$), triangular ($n=3$), rectangular ($n=4$), etc.
distortions of the jet.  In cylindrical geometry $n$ is an integer and
represents  the azimuthal wavenumber.  For $n>0$, waves propagate at an
angle to the flow direction, and $+n$ and $-n$ specify wave propagation
in the clockwise and counterclockwise sense, respectively, when viewed
outwards along the flow direction.

\newpage
\subsection{The Low Frequency Limit}

Each wave mode {\it n} contains a {\it surface} wave and multiple {\it
body} wave solutions to the dispersion relation. In the low frequency
limit the real part of the pinch mode ($n=0$) {\it surface} wave
solution becomes (Hardee 1995)
\begin{equation}
\label{1}\frac \omega k\approx u\pm \left\{ \frac 12\left( V_{A,jt}^2+\frac{%
V_{A,jt}^2a_{jt}^2}{(a_{jt}^{ms})^2}\right) \pm \frac 12\left[ \left(
V_{A,jt}^2+\frac{V_{A,jt}^2a_{jt}^2}{(a_{jt}^{ms})^2}\right) ^2-4\frac{%
V_{A,jt}^4a_{jt}^2}{(a_{jt}^{ms})^2}\right] ^{1/2}\right\} ^{1/2} \text{ ,}
\end{equation}
where $a_{jt}^{ms} \equiv \left( a_{jt}^2+V_{A,jt}^2\right) ^{1/2}$ is
the magnetosonic speed, $a_{jt}=(\Gamma p_{jt}/\rho _{jt})^{1/2}$ is
the jet sound speed, and $V_{A,jt}=(B_{jt}^2/4\pi \rho _{jt})^{1/2}$ is
the jet Alfv\'en speed. The imaginary part of the solution is
vanishingly small in the low frequency limit. These solutions are
related to fast ($+$) and slow ($-$) magnetosonic waves propagating
with ($u+$) and against ($u-$) the jet flow speed $u$, but strongly
modified by the jet-external medium interface.  The unstable growing
solution is associated with the backwards moving (in the jet fluid
reference frame) slow magnetosonic wave and in the supermagnetosonic
limit moves at nearly the flow speed in the observer frame.

In this limit all higher order modes ($n>0$) have {\it surface} wave
solutions given by
\begin{equation}
\label{2}\frac{ku}\omega \approx \frac 1{1-V_{A,jt}^2/u^2}\left\{ 1\pm
i\left[ \left( 1-\frac{V_{A,jt}^2}{u^2}\right) \frac 1\eta -\frac{V_{A,jt}^2%
}{u^2}\right] ^{1/2}\right\} \text{ ,} 
\end{equation}
where $\eta \equiv \rho _{jt}/\rho _{ex}$. Spatial growth corresponds
to the minus sign in equation (2) and a negative value for the
imaginary part of the complex wavenumber. In the dense jet limit, i.e.,
$\eta \rightarrow \infty $, equation (2) becomes $\omega /k\approx u\pm
V_{A,jt}$ , and the surface waves are related to Alfv\'en waves
propagating with and against the jet flow speed u, but strongly
modified by the jet-external medium interface. The unstable growing
solution is associated with the backwards moving (in the jet fluid
reference frame) Alfv\'en wave. In the supermagnetosonic limit the real
part of $\omega /k\approx [\eta /(1+\eta)]u$ in the observer frame and
the wave speed is a strong function of the density ratio.

In this limit the {\it body} wave solutions are purely real, non-propagating 
and given by
\begin{equation}
\label{3}kR=\frac{(n+2m-1/2)\pi /2}{\left[ \frac{(M_{jt}^{ms})^2}{%
1-(M_{jt}^{ms}/M_{jt}M_{jt}^A)^2}-1\right] ^{1/2}} \text{ ,} 
\end{equation}
where {\it n} is the mode number, and $m\geq 1$ is the integer body
mode number. Additionally, $M_{jt}\equiv u/a_{jt}$ is the sonic Mach
number, $M_{jt}^A\equiv u/V_{A,jt}$ is the Alfv\'en Mach number, and
$M_{jt}^{ms}\equiv u/a_{jt}^{ms}$ is the magnetosonic Mach number.

\newpage
\subsection{The Maximum Growth Rate}

With the exception of the pinch mode surface wave which has a
relatively broad plateau in the growth rate, all other solutions have a
relatively sharp maximum in the growth rate when a jet is
supermagnetosonic. In the supermagnetosonic limit, a maximum in the
growth rate is achieved for
\begin{equation}
\eqnum{4a}\omega R/a_{ex}\approx \omega _{nm}^{*}R/a_{ex}\equiv
(n+2m+1/2)\pi /4\text{ , } 
\end{equation}
\begin{equation}
\eqnum{4b}\lambda \approx \lambda _{nm}^{*}\equiv \frac{2\pi }{\omega
_{nm}^{*}R/a_{ex}}\frac{M_{jt}^{ms}}{1+M_{jt}^{ms}/M_{ex}}R\text{ ,} 
\end{equation}
\begin{equation}
\eqnum{4c}v_w\approx v_w^{*}\equiv \frac{M_{jt}^{ms}/M_{ex}}{%
1+M_{jt}^{ms}/M_{ex}}u\text{ ,} 
\end{equation}
\setcounter{equation}{4}
where {\it n} gives the wave mode, $m=0$ gives the surface wave, $m\geq
1$ gives the body waves. The wave speed at the maximum growth rate is
significantly different from the low frequency wave speed. 

With the exception of the $n=0$, $m=0$, pinch mode surface wave, the
maximum spatial growth rate is approximately given by
\begin{equation}
\label{5}k_I\approx k_I^{*}\equiv -(2M_{jt}^{ms}R)^{-1}\ln \left[ 4\frac{%
\omega _{nm}^{*}R}{a_{ex}}\right] \text{ ,} 
\end{equation}
where $k_I$ is the imaginary part of the complex wavenumber. When a jet
is transmagnetosonic but super-Alfv\'enic, the growth rate of the
helical and higher order surface waves has no maximum but increases
proportional to the frequency, and the growth rate of the body waves
decreases. When a jet is sub-Alfv\'enic, the helical and higher order
surface waves are stable (Bodo et al. 1989; Hardee et al. 1992) but the
pinch surface wave can remain unstable. Body waves exist and are
unstable provided the denominator of equation (3) is real, and this
occurs if the jet speed exceeds the fast magnetosonic speed, or if the
jet speed is slightly below the slow magnetosonic speed (Bodo et al.
1989; Hardee et al. 1992).

\subsection{The High Frequency Limit}

In the high frequency limit the real part of the solutions to the
dispersion relation for all surface and body waves is given by
\begin{equation}
\label{6}\frac \omega k\approx u\pm \left[ \frac
12(a_{jt}^2+V_{A,jt}^2)\left\{ 1\pm \left[ 1-\frac{4a_{jt}^2V_{A,jt}^2}{%
(a_{jt}^2+V_{A,jt}^2)^2}\right] ^{1/2}\right\} \right] ^{1/2}\text{ .} 
\end{equation}
Equation (6) describes fast, $\omega /k\approx u\pm
{\rm Max}(V_{A,jt},a_{jt})$, and slow, $\omega /k\approx u\pm
{\rm Min}(V_{A,jt},a_{jt})$, magnetosonic waves.  The unstable growing
solution is associated with the backwards moving (in the jet fluid
reference frame) fast magnetosonic wave but the growth rate is
vanishingly small in the high frequency limit.  The wave speed in the
observer frame is again considerably different from the wave speed at
low frequency and at the maximum growth rate.

\subsection{Numerical Solutions}

We have solved the dispersion relation numerically using root finding
techniques over a wide range of perturbation frequencies for parameters
appropriate to the three numerical simulations. Figure 12 shows the
numerical solution for the surface and first three body waves
associated with the helical ($n=1$), elliptical ($n=2$), triangular
($n=3$) and rectangular ($n=4$) modes for parameters appropriate to the
heavy jet in simulation C, and Table 2 contains the wavelengths at the
maximum growth rate. The body waves have a larger maximum growth rate
than the surface wave only for the pinch (not shown) and helical modes.
We note that the maximum growth rate of the $n>1$ surface wave modes
exceeds the maximum growth rate given by equation (5) but that the
computed maximum growth rate of {\it all} body wave modes remains below
the maximum predicted by equation (5). The vertical line in the panels
is at the precessional perturbation frequency used in simulation C.
Numerical solutions for parameters appropriate to the other two
simulations are qualitatively similar. Quantitative differences between
numerical solution of the dispersion relation for simulations A and B
occur largely in the maximum growth rate as a result of the slightly
different magnetosonic Mach numbers and in the wavelength as a function
of frequency as a result of different wave propagation speeds for the
different density jets.

The displacement and velocity perturbations associated with the normal modes
of a cylindrical jet can be written in the form (see Appendix)
\begin{equation}
\eqnum{7a} \text{\boldmath$ \xi$}(r,\phi _s,x_s)={\bf A}(r)e^{i{\bf \Delta }(r)}\xi
_{r,n}(R)\exp [i(kx_s\pm n\phi _s-\omega t)]\text{ ,} 
\end{equation}
\begin{equation}
\eqnum{7b}{\bf u}_1=-i(\omega-ku) {\bf A}(r)e^{i{\bf \Delta }(r)}\xi
_{r,n}(R)\exp [i(kx_s\pm n\phi _s-\omega t)]\text{ ,}
\end{equation}
\setcounter{equation}{7}
where  {\bf u}$ _1(r,\phi ,x)=d${\boldmath$ \xi$}$/dt$, $\xi _{r,n}(R)$
is the surface amplitude of the radial displacement, $x_s$ is the
unperturbed axial location at the jet surface, $\phi _s$ is the
azimuthal phase angle at the jet surface, and {\bf A}(r) $\leq 1$ and
${\bf \Delta}$(r) give the amplitude and phase dependence relative to
the jet surface, respectively. If the dependence of the radial fluid
displacement inside the jet on rotation in azimuth is ignored then the
radial fluid displacement of a surface wave mode {\it n} is
approximately given by $\xi _r(r)\approx \left[\xi
_{r,n}(R)\right](r/R)^{n-1}$ and therefore displacements produced by
higher order surface modes are predicted to show a much more rapid
decrease inwards (Hardee 1983). In fact the surface waves show a
somewhat faster fall off in amplitude relative to the surface amplitude
than that predicted by the analytical approximation as a result of
significant rotation in azimuth of, say, the maximum internal
displacement relative to the maximum surface displacement (Hardee,
Clarke, \& Howell 1995). At a constant azimuth the body waves show a
reversal in fluid displacement at null surfaces interior to the jet
surface. Only the helical surface and body waves induce displacement of
the jet center.  Displacements produced by body waves are also, in
general, largest at the jet surface and decrease inwards (see Hardee,
Clarke, \& Howell 1995).

Fluid displacement surfaces associated with the helical, elliptical,
triangular, and rectangular mode surface and first three body waves
computed from equation (7a) for parameters appropriate to the heavy jet
simulation C at the appropriate maximally growing frequencies (see
Figure 12) are shown in Figure 13. Note how the distortion of inner
surfaces becomes more out of phase with the jet surface as the jet
center is approached for all surface and body modes but that little
internal distortion accompanies the triangular and rectangular surface
wave modes. Behavior appropriate to the simulations is primarily of the
form $f_1(r,\phi ,x)=A(r)e^{i\Delta (r)}f_1(R)\exp [i(kx_s-n\phi
_s-\omega t)]$ and over time ($x_s={\rm const.}$, $n\phi _s+\omega
t={\rm const.}$) patterns rotate in the counterclockwise direction
(flow is into the page as in the magnetic pressure cross sections
associated with the simulations). Thus, internal displacement surfaces
exhibit a phase lead relative to the surface displacement, e.g., the
innermost displacement surface shown associated with the helical
surface and first body waves leads the surface displacement by about
30$^{\arcdeg}$ and 225$^{\arcdeg}$, respectively. The spiral patterns
associated with the displacement surfaces are ``trailing'' as the
patterns rotate in the counterclockwise direction at fixed position.
Note, however, that as one moves out along the jet axis ( $t={\rm
const.}$, $k_Rx_s-n\phi _s={\rm const.}$) all patterns twist in a
clockwise sense with a spatial rate $\phi _s^n(x)=(k_R/n)x_s$, where
$k_R$ is the real part of the wavenumber.

In Figure 13, approximately the largest possible displacements of
initially circular surfaces are shown to emphasize the distortions, and
a maximum in the surface displacement is chosen to lie along the
positive horizontal axis ($z$-axis in the simulations). At least
approximately, the maximum displacement amplitude must be less than
that leading to the overlap of displacement surfaces computed from the
linearized equations because overlap implies the development of shocks.
Using this criterion we find that the maximum amplitude associated with
the maximum growth rate for the surface wave solutions to the helical,
elliptical, triangular and rectangular modes is given by
\begin{equation}
\label{8}\left[ \xi _{r,n}^{*}(R)\right] ^{\max }\approx \frac{0.9}{n+1/2}R
\text{ ,} 
\end{equation}
and for the first three body wave solutions to the pinch, helical,
elliptical, triangular and rectangular modes is given by
\begin{equation}
\label{9}\left[ \xi _{r,nm}^{*}(R)\right] ^{\max }\approx \frac{0.7}{
n+2m+1/2}R\text{ .} 
\end{equation}
The maximum amplitudes are basically $\left[ \xi _{r,nm}^{*}\right]
^{\max }\propto \left( \omega _{nm}^{*}\right) ^{-1}\propto \left(
n+2m+1/2\right) ^{-1}$, and maximum displacements become smaller for
higher order surface waves (larger $n$) and higher order body waves
(larger $m$). For longer or shorter wavelengths, i.e., waves at
frequencies below or above maximum growth, larger or smaller maximum
surface amplitudes are allowed with the maximum allowed amplitude
\begin{equation}
\label{10}\xi _{r,nm}^{\max }(R)\approx (\lambda _{nm}/\lambda
_{nm}^{*})(v_w/v_w^{*})\left[ \xi _{r,nm}^{*}(R)\right] ^{\max }\text{,} 
\end{equation}
where $\lambda _{nm}^{*}$ and $v_w^{*}$ are the wavelength and wave
speed at the maximum growth rate. These surface amplitudes require
velocity perturbations transverse to the flow less than the internal
jet magnetosonic speed, and lesser amplitudes require velocity
perturbations $u_{r1}(R)/$ $u_{r1}^{\max }(R)\approx \xi _{r,nm}(R)/\xi
_{r,nm}^{\max }(R)$. At the maximum amplitudes the perturbations are no
longer linear.

\newpage
\section{Comparison with Simulations}

The predicted stability properties of jets in the supermagnetosonic
regime are (1) a very small growth rate for the pinch mode surface
wave, (2) a slightly higher maximum growth rate for all the body wave
modes compared to the maximum growth rate of the helical mode surface
wave, and (3) a higher maximum growth rate for higher order surface
waves compared to the maximum growth rate of all the body wave modes.
The computed growth rates along with the maximum allowed displacements
would lead us to predict that significant dynamical structures will be
produced by a few of the lower order surface and body wave modes. Of
these few modes the higher order modes should achieve maximum
displacements closer to the jet inlet as a result of the faster spatial
growth rate, and the lower order modes should achieve larger maximum
displacements farther from the jet inlet. Note that in the simulations
the inlet perturbation favors development of the helical surface mode
and thus higher order body and surface modes may not develop as close
to the inlet as might be expected if all wave modes were excited with
the same initial amplitude. Nevertheless, in the simulations we see
that the higher order modes develop first to some maximum amplitude and
then decline in amplitude as the lower order modes grow to larger
amplitudes. In simulations A and B the larger amplitudes associated
with the helical and elliptical modes lead to disruption of organized
structure. We note that in each simulation, the propagation speed of
surface and body wave modes is similar and is within 10\% of the value
computed from equation (4c).  Interestingly, the wave speeds observed
in the simulations $\left[{\it computed~from~the~linear~theory}\right]$
are 0.30u $\left[ {\it 0.32u}\right]$ from simulation A, 0.54u $\left[
{\it 0.49u}\right]$ from  simulation B, and  0.74u $\left[ {\it 0.66u}
\right]$ from simulation C.  This reveals a definite trend towards
observed speeds being higher than computed speeds as the jet density
increases.

All simulations show similar structures in the linear limit. Because
the structures are better defined in simulation C and grow to nonlinear
amplitudes without disruption, we turn to comparison with simulation C
to obtain nonlinear estimates of the maximum amplitudes associated with
surface and body wave modes. In order to identify the surface modes
responsible for structures observed in the numerical simulations, we
have attempted to fit the well-defined jet surface distortions observed
in cross sections in simulation C using displacement surfaces computed
from equation (7a) and the results are shown in Figure 14.  The
displacement surfaces shown in Figure 14 were produced as a linear sum
of displacements produced by helical, elliptical, triangular and
rectangular surface waves at the positions of the cross sections shown
in Figure 4 at time $\tau_d =34$.  No fit was attempted at an axial
distance of 48$R$ as the jet had not yet reached a quasi-equilibrium
configuration at this distance by this dynamical time. The cross
sections can be fit by: (1) A helical distortion whose phase angle
rotates through 360$^{\arcdeg}$ over a wavelength $\lambda _1\approx
14R$ and which builds to a maximum surface amplitude of $\xi
_{r1}^{obs}/R \approx 0.8$ at an axial distance of 36$R$ and
maintains this amplitude at larger distance. (2) An elliptical
distortion that rotates through a phase angle of 180$^{\arcdeg}$ over a
wavelength $\lambda _2\approx 14R$ and which builds to a maximum
surface amplitude of $\xi _{r2}^{obs}/R\approx 0.6$ at an axial
distance of 42R and maintains this amplitude at larger distance. (3) A
triangular distortion whose phase angle rotates through 120$^{\arcdeg}$
over a wavelength $\lambda _3\approx 9R$, and whose amplitude builds to
a maximum amplitude of $\xi _{r3}^{obs}/R\approx $ $0.1$ at an axial
distance $\approx 18R$ and declines in amplitude at larger axial
distance. (4) A rectangular distortion which rotates through a phase
angle of 90$^{\arcdeg}$ over a wavelength $\lambda _4\approx 6R$ as
suggested by animation of the cross section images from the numerical
simulation. The simulation animations indicate the presence of both
$+n$ (clockwise rotating) and $-n$ (counterclockwise rotating) surface
waves of equal amplitude which reinforce the square shape in
45$^{\arcdeg}$ increments with a multiply fluted cross sectional
appearance at intermediate angles.  The counter rotating waves
reinforce in a linear additive fashion and grow to a  maximum amplitude
of $\xi _{r4}^{obs}/R\approx 0.07$ (0.14 combined) at an axial
distance $\approx 18R$ and decline in amplitude at larger axial
distance.

The wavelengths used to fit the observed surface structures lie midway
between the fastest growing wavelength and a precessionally induced
wavelength for the helical mode, i.e., $12.5R < \lambda _1 < 16.3R$,
and at a wavelength 50\%, 40\% and 20\% longer than the fastest growing
wavelength for the elliptical, triangular and rectangular modes,
respectively.  This result probably indicates a stronger coupling
between the initial perturbation frequency and the helical mode and
weaker coupling to the higher order modes. The surface displacements
observed in the simulation have been fit reasonably well using
displacement surfaces computed from the linear theory even at the
nonlinear amplitudes observed for the helical and elliptical
distortions. The observed wavelength and the maximum observed surface
displacements compared to the predicted maxima (computed from eqs. 8 \&
10) are shown in Table 3.  The maximum observed surface displacements
for helical and elliptical modes are approximately equal to the
computed maxima.  The maximum observed surface displacements for
triangular and rectangular modes are less than the computed maxima. The
transverse velocity perturbation as computed from equation (7b)
required to accompany the observed displacements is less than the jet
magnetosonic speed for all observed surface modes.

We have performed a preliminary investigation of the velocity structure
of surface and body waves along the jet axis in an attempt to identify
the periodicities observed in the velocity plots with either surface or
body waves. In particular, let us concentrate on the velocity plots
shown in Figure 8 for simulation C. Note that a multiplication by a
factor 1.9 converts the velocity scale in units of $a_{ex}$ in Figure 8
to multiples of the jet magnetosonic speed, $a_{jt}^{ms}$. Refer to
Table 2 for the wavelengths of surface and body waves at maximum growth
computed from the linear analysis. Interior to an axial distance of
30$R$ velocity fluctuation in $u_x$ ($x$-velocity panel in Figure 8)
shows scale lengths, $1.5-3.5R$, that lie in the range of the first
three pinch mode body waves. In general, in the linear limit the higher
order modes can produce only relatively small velocity fluctuation in
the axial direction.  On the other hand, the large dips in $u_x$ at
axial distances of $\sim 32R$ and $\sim 48R$ and accompanying
transverse velocity maxima in $u_y$ and $u_z$ ($y$- and $z$-velocity
panels in Figure 8) are the result of a combination of large amplitude
helical and elliptical jet perturbation.  Basically the jet at these
locations has been displaced off the axis (see, for example, Figure 5
and Figure 14) with axial velocity representative of the jet sheath.

The readily identifiable structures in the transverse velocity panels
in Figure 8 are transverse velocity oscillation associated with the
large scale helical twist ($\lambda \approx 14R$), a relatively large
transverse velocity oscillation at considerably smaller wavelength
($\lambda \approx 3R$) that can be identified with the first helical
body wave, and still shorter wavelength ($\lambda \approx 1.5R$)
transverse velocity oscillation structure in the ranges of the second
and third helical body waves. We note that the transverse velocities
observed along the jet axis, which can largely be attributed to the
helical mode, are typically less than the jet magnetosonic speed
(recall that $a_{jt}^{ms}\sim 0.5a_{ex}$)  We note that the
line-of-sight images showed dual twisted filaments internal to the jet
($\lambda \approx 4R$) which suggest the presence of internal
elliptical body waves at axial distances $<30R$. No significant
transverse velocity structure along the jet axis is predicted to
accompany this wave mode or with higher order wave modes.  The observed
body mode structures are not far above the limits imposed by the
computational grid. It is likely that the amplitudes of these
structures are influenced by the grid scale.  A comprehensive
investigation of internal jet structure and velocity awaits the
completion and analysis of considerably higher resolution numerical
simulations (Hardee et al., in preparation).

\section{Summary and Discussion}

A linearized analysis of supermagnetosonic jet stability predicts that
the helical, elliptical, triangular and rectangular surface wave modes
can significantly influence jet dynamics and structure. Additionally,
the first few body waves associated with pinch, helical, elliptical,
triangular and rectangular modes are also predicted to be capable of
significantly influencing jet dynamics and structure. In the present
simulations the line-of-sight and cross section images show compelling
evidence for helical, elliptical, triangular and rectangular {\it
surface} distortions and provide some evidence for pinch, helical and
elliptical {\it body} distortions. Helical and elliptical distortions
are seen in the simulations at up to the maximum amplitudes predicted
by the linear stability analysis. A rapid decline in the maximum
amplitudes of waves for higher order modes is predicted by the linear
analysis. However, a more rapid decline than predicted in the maximum
amplitudes of the higher order {\it surface} wave modes is observed in
the simulations, e.g., triangular and rectangular {\it surface} wave
modes are seen at much less than the maximum amplitudes allowed by the
linear theory.  It is possible that the amplitudes of the observed
higher order modes have been artificially reduced by the grid scale
used in the simulations. Verification of this rapid decline in
amplitudes for the higher order {\it surface} modes awaits completion
and analysis of considerably higher resolution numerical simulations.
Comparison between the present numerical simulations and a linear
stability analysis shows that the linear analysis does a remarkable job
of fitting jet structures even at the largest amplitudes observed.

Velocity structure in the numerical simulations has been compared to
velocity structure predicted by the linear analysis to accompany {\it
surface} wave modes.  The velocity structure in the jet is consistent
with the amplitudes of the {\it surface} distortions observed in the
simulations. Velocity structure reveals the presence of jet {\it body}
distortions that are not apparent in the line-of-sight images.  In
these simulations, the small transverse velocity supplied by precession
at the jet origin, lack of jet rotation and weak toroidal magnetic
field provide a very small angular momentum.  All wave modes other than
the pinch mode consist of counterclockwise (sense of jet precession and
of magnetic helicity) and clockwise growing components.  The sense of
jet precession and of magnetic helicity provide only a small bias
towards the counterclockwise rotating components.  In fact, nearly
equal amplitude counter rotating components were observed for the
rectangular mode.  At a much lesser level this effect may be present in
the helical and elliptical modes but is obscured by the combined
distortions.

All three jets show evidence for similar structure before helical and
elliptical distortions grow to large amplitudes.   Organized helical and elliptical structures persist at large amplitudes only for the jet
denser than the external medium. For example, the jet less dense
than the external medium loses structure and becomes plume-like as
helical and elliptical distortions reach large amplitude and the jet
entrains denser external material. Conversely, the dense jet maintains
a high speed spine surrounded by a low speed sheath across the
computational grid. A jet of density equal to that of the external
medium is intermediate between these two cases and shows some evidence
for a spine-sheath morphology although the spine slows significantly on
the computational grid. When compared to previous purely fluid fully
three-dimensional simulations, our present simulation results show that
the inclusion of a relatively strong, equipartition, nearly poloidal
magnetic field does not significantly modify the linear structure or
the nonlinear entrainment and slowing of a jet less dense than the
external medium found by Hardee, Clarke, \& Howell (1995).  Note also
that the present simulations contain magnetic fields that are somewhat
stronger than might be expected in an astrophysical jet if a large lobe
is to form (Clarke 1994), and contain no magnetic field in the external
medium as might be expected within a lobe or cocoon formed from jet
material.  We do not expect reduction in the jet magnetic field or
inclusion of a weak external magnetic field to  change the present
results substantially but additional simulations designed to study in
more detail the effect of magnetic field orientation on structure,
entrainment and mixing are in progress.

Although the magnetosonic Mach numbers in the three simulations are
comparable, the jet velocity is 12.5 $a_{ex}$, 5.14 $a_{ex}$ and 4.04
$a_{ex}$ in the light jet, equal density jet and heavy jet simulations,
respectively.  Previous purely fluid light jet simulations showed that
higher jet velocity leads to more rapid heating and disruption as jet
and external fluids mix.  Thus, in these simulations it is likely that
the rapid disruption of the light jet and lack of disruption of the
heavy jet is at least partially attributable to the velocity difference
and not just the density difference.  However, note that the most
stable combination (linearly and nonlinearly) for a given magnetosonic
Mach number is always associated with the denser jet.  This is so
because for a fixed magnetosonic Mach number, which governs the spatial
growth of perturbations in the linear regime, the denser jet will have
a lower velocity relative to the external medium than the light jet and
hence be less susceptible to nonlinear disruption.

The jets in extragalactic radio sources can extend to distances that
are considerably larger in terms of the initial jet radius than the
distances studied in these numerical simulations.  Transverse
perturbations to the jet may be in the form of the regular precession
studied here, although at considerably different frequency, or may be
more random in nature.  That small transverse perturbations can
propagate to very large distances has been demonstrated in
two-dimensional slab jet simulations.  A slab jet is spatially resolved
along two Cartesian axes and is effectively infinite in extent in the
third dimension.  A set of simulations of expanding ``equilibrium''
slab jets in suitable atmospheres has shown propagation and growth of
sinusoidal oscillations out to distances over two hundred times the
initial jet radius (Hardee \& Clarke 1995).  In these simulations
adiabatic jet expansion was shown to provide a predicted stabilizing
influence through both increase in the local Mach number and increase
in the fundamental length scale given by the local jet radius.  More
recent simulations of constant radius ``equilibrium'' slab jets in a
constant atmosphere but with very high Mach numbers has shown
propagation and growth of sinusoidal oscillations out to distances in
excess of six hundred times the jet radius (Stone, Xu, \& Hardee
1997).  These slab jet simulations make it clear that small linear
disturbances initiated near the origin of a jet can lead to
consequences at much larger scales and thus our present studies at
relatively small spatial scales can be imagined to apply to much larger
spatial scales.

Random transverse perturbations can be initiated at any point along a
long jet whereas periodic perturbations are almost certainly produced
near to the origin.  Previous slab jet simulations have shown that a
random perturbation excites wave mode development at the most unstable
frequency and wavelength (Zhao et al.\ 1992).  With no perturbation at
the jet origin, a three-dimensional propagating jet simulation shows
essentially no instability until random perturbations driven by
asymmetric lobe turbulence excite wave mode development at the most
unstable frequencies and wavelengths (Norman 1996).  Previous numerical
simulations of (slab) cylindrical jets undergoing  periodic
(oscillation) precession show that excitation of all wave modes other
than the large scale (sinusoid) helix occurs at or near the most
unstable frequency and wavelength (cf. the higher order surface and
body modes in these simulations).  Periodic perturbations are not
likely to be at the most unstable helical frequency as was the case in
the present numerical simulations.  However, it has been shown that low
precession frequencies result in behavior like that seen in these
simulations albeit with the longer helical wavelength and slower
spatial growth predicted by the linear stability analysis; precession
frequencies above the most unstable helical frequency should only be
capable of exciting internal structures as the jet cannot respond
bodily to higher frequencies (Hardee, Cooper, \& Clarke 1994; Stone,
Xu, \& Hardee 1997).  Since lower frequency precessional perturbations
result in some stabilization as helical twisting is of longer
wavelength and grows more slowly, somewhat slower development of the
nonlinear behavior (primarily the result of helical and elliptical
distortion) observed in these simulations is likely to be the most
significant difference.  Note that high frequency precession may also
lead to increased stability as large scale helical twisting should be
suppressed. This situation is sufficiently different from our present
simulations that our present results should not be extended to the
situation of rapid precession.  Our results should also not be extended
to the case of large amplitude nonlinear initial perturbations.

Magnetic jet acceleration and collimation schemes require and simple
flux conservation arguments combined with jet expansion imply that
magnetic fields in large scale jets take on a primarily toroidal
configuration different from the poloidal configuration studied here.
Preliminary results indicate that weak toroidal and/or poloidal fields,
i.e., plasma $\beta \equiv 8 \pi P_{th}/B^2 >> 1$, will behave like the
present simulations. Only strong toroidal magnetic fields will make a
large difference in jet and radio source morphology.  Note also that
numerical simulations show that strong toroidal magnetic fields inhibit
the formation of a large radio lobe (Clarke 1994) suggesting that
strong toroidal fields are not present at large scales.  Thus, we are
confident that the present poloidally magnetized numerical simulations
serve as a reliable indicator of the behavior of many extragalactic jet
outflows on the largest scales.

The nonlinear behavior and appearance of the jets in these simulations
suggests that FR~I radio sources have a plume-like appearance, e.g.,
M\,84, or develop a plume-like appearance after exhibiting large scale
helical distortion, e.g., 3C\,449 (see Hardee, Cooper, \& Clarke
1994), as a result of jet interaction with a medium that is of
comparable or greater density than the jet.  This happens if the
propagating jet does not form or cannot maintain a low density lobe or
cocoon that can protect the jet from a denser external environment.
Note that formation and maintenance of a lobe or cocoon requires both a
low density jet with respect to the undisturbed external environment
and sufficiently high pressure at the jet front to inflate a lobe and
drive a backflow against the expected pressure gradients associated
with a galactic atmosphere (cf. Hardee et al.\ 1992). On the other
hand, these simulations suggest that FR~II radio sources can have long
structured jets because the jet forms and maintains a lobe or cocoon of
density much less than that of the jet through which the jet propagates
without disruptive entrainment.

The present dense jet simulation shows that helical twisting and
twisted elliptical jet distortion, which bifurcates the poloidal
magnetic field into filaments, can exist over large distances without
leading to rapid disruption of organized jet structures. Additionally,
the elliptical distortion and accompanying magnetic filaments can twist
at a rate much different from the helical twist in a dense jet (unlike
the light jet). This jet morphology appears remarkably similar to the
morphology of the Cygnus A jet (Carilli et al.\ 1996), which shows
evidence for a filament pair that twists through 180$^{\arcdeg}$ over a
distance comparable to that of the 360$^{\arcdeg}$ helical twist
suggested by the sinusoidal oscillation of the jet in the plane of the
sky, and polarization vectors suggesting primarily poloidal magnetic
fields. Additionally, the dense jet simulation shows that a magnetic
filament brightens when it is aligned with the leading edge of jet
helicity, which leads to bright diagonal filaments that cross the jet
with the same angle at regular intervals.  Similar diagonal structure
is also evident in the Cygnus A jet.  Analysis of the structures in the
Cygnus A jet indicates a jet with a relativistic Mach number $M_{rel}
\equiv \gamma_{rel}/\gamma_s(u/a_{jt}) \approx 10$ (Hardee 1996) where
$\gamma_{rel}$ and $\gamma_s$ are the Lorentz factors corresponding to
the jet flow speed and sound speed, respectively.

In the dense jet simulation the jet as revealed by the total intensity
image exhibits a relatively large-amplitude transverse oscillation.
This simulation shows that the jet consists of a small jet core
embedded in a broader sheath.  Thus, a modest transverse motion
relative to the width of the core and sheath could appear as a
relatively large displacement of the observed jet core.  We note that
the sheath in the present dense jet simulation extends somewhat beyond
the region into which jet fluid and magnetic field have been mixed with
the external medium. We cannot draw any firm conclusions regarding the
transverse extent of a sheath in extragalactic jets from these
simulations as numerical effects and the discrete grid lead to an
effective Reynolds number different from that of an extragalactic jet.
However, some of the qualitative aspects of our present simulations
should hold true for extragalactic jets and we should not be surprised
if the highly collimated outflows seen in continuum radio images are
embedded in more slowly moving outwards flowing material which may also
exhibit helically twisted magnetized emission filaments at some
distance from the jet core.

\acknowledgements
P. Hardee and A. Rosen acknowledge support from the National Science
Foundation through grant AST-9318397 to the University of Alabama. D. Clarke
is supported in part by the Natural Sciences and Engineering Research
Council of Canada (NSERC). The numerical work utilized the Cray C90 at the
Pittsburgh Supercomputing Center through grant AST930010P.

\newpage 

\appendix 

\section{Appendix}

Let us model a jet as a cylinder of radius R, having a uniform density,
$\rho _{jt}$, a uniform internal poloidal magnetic field, B$_{jt}$, and
a uniform velocity, $u$. The external medium is assumed to have a uniform
density, $\rho _{ex}$, and to contain no magnetic field. The jet is
established in static total pressure balance with the external medium
where the total static uniform pressure is $p_{jt}^{*}\equiv
p_{jt}+B_{jt}^2/8\pi=p_{ex}^{*}=p_{ex}$. The general approach to
analyzing the stability properties of this system is to linearize the
one-fluid MHD equations of continuity and momentum along with an
equation of state within each medium where the flow velocity $u = 0$. The
flow velocity is then reintroduced when solutions are matched at the
jet boundary. In each medium we assume that the perturbations are
adiabatic in nature. The linearized ideal magnetohydrodynamic equations
that are relevant to our model become
$$
\frac{\partial \rho _1}{\partial t}+{\bf \bigtriangledown }\cdot (\rho _0
{\bf u}_1)=0 \text{ ,} 
$$
$$
\rho _0\frac{\partial {\bf u}_1}{\partial t}=-{\bf \bigtriangledown }p_1+
\frac{\left[ (\bigtriangledown \times {\bf B}_0)\times {\bf B}
_1+(\bigtriangledown \times {\bf B}_1)\times {\bf B}_0\right] }{4\pi } \text{ ,}
$$
$$
\frac{\partial p_1}{\partial t}=-\Gamma p_0({\bf \bigtriangledown }\cdot 
{\bf u}_1) \text{ ,} 
$$
where the density, velocity, pressure, and magnetic field are written
as $\rho =\rho _0+\rho _1$, ${\bf u}={\bf u}_1$\text{,}
$p=p_0+p_1$, ${\bf B}={\bf B}_0+{\bf B}_1$, and subscript 1 refers to
a perturbation to the equilibrium quantity with subscript 0.

In cylindrical geometry we look for perturbations $\rho _1$,{\bf
\ }${\bf u} _{1\text{,}}$ $p_1$, and ${\bf B}_1$ of the form
\begin{equation}
\label{A1}f_1(r,\phi ,x)=f_1(r)\exp [i(kx\pm n\phi -\omega t)] \text{ ,}
\end{equation}
where the flow is in the positive $x$ direction, and $r$ is in the radial
direction with the flow bounded by $r=R$. With this form for the
perturbations the differential equation for the dependence of the total
pressure perturbation $ p_1^{*}=p_1+({\bf B}_1\cdot {\bf B}_0)/4\pi $
within each fluid as a function of $r$ can be written in the form
\begin{equation}
\label{A2}r^2\frac{\partial ^2}{\partial r^2}p_1^{*}+r\frac \partial
{\partial r}p_1^{*}+\left[ \beta ^2r^2-n^2\right] p_1^{*}=0\text{ ,} 
\end{equation}
where for poloidal magnetic fields
\begin{equation}
\label{A3}\beta \equiv \left[ -k^2+\frac{\omega ^4}{\omega
^2(a^2+V_A^2)-k^2V_A^2a^2}\right] ^{1/2}\text{ .} 
\end{equation}
In equation (A3), $a=\left( \Gamma p/\rho \right) ^{1/2}$ is the sound
speed, and $V_A=\left( B^2/4\pi \rho \right) ^{1/2}$ is the Alfv\'en
speed in the appropriate medium. The solutions that are well behaved at
jet center and at infinity are
\begin{equation}
\label{A4}p_{1jt}^{*}(r)=C_{jt}J_{\pm n}(\beta _{jt}r)\text{ , and }
p_{1ex}^{*}(r)=C_{ex}H_{\pm n}^{(1)}(\beta _{ex}r) 
\end{equation}
inside and outside the jet, respectively, where $J_{\pm n}$ and $H_{\pm
n}^{(1)}$ are Bessel and Hankel functions. In equation (A4)
$$
\beta _{jt}=\left[ -k^2+\frac{(\omega -ku)^4}{(\omega
-ku)^2(a_{jt}^2+V_{A,jt}^2)-k^2V_{A,jt}^2a_{jt}^2}\right] ^{1/2}\text{, and }
\beta _{ex}=\left[ -k^2+\frac{\omega ^2}{a_{ex}^2}\right] ^{1/2}\text{,} 
$$
there is no magnetic field in the external medium, and $\omega
\rightarrow \omega -ku$ has reintroduced the jet flow speed through the
Doppler shifted frequency in $\beta _{jt}$.

At $r = R$ we require that the total pressure be continuous across the
boundary, i.e., $p_{1jt}^{*}(R)=p_{1ex}^{*}(R)$, and that the radial
fluid displacement inside and outside the jet be equal at the jet
boundary, i.e., $ \xi _r^{jt}(R)=\xi _r^{ex}(R)$, where the fluid
displacement in the radial direction, $\xi _r(r)$, is given by
\begin{equation}
\label{A5}\xi _r^{jt}(r)=\frac 1{\chi _{jt}}\frac{\partial p_{1jt}^{*}(r)}{
\partial r}\text{ , and }\xi _r^{ex}(r)=\frac 1{\chi _{ex}}\frac{\partial
p_{1ex}^{*}(r)}{\partial r} 
\end{equation}
inside and outside the jet, respectively, and where
$$
\chi _{jt}=\rho _{jt}[(\omega -ku)^2-k^2V_{A,jt}^2]\text{ , and }\chi
_{ex}=\rho _{ex}\omega ^2. 
$$
Equations (A4) and (A5) along with the boundary conditions on the
pressure and radial displacement result in a dispersion relation
\begin{equation}
\label{A6}\frac{\beta _{jt}}{\chi _{jt}}\frac{J_n^{^{\prime }}(\beta _{jt}R)%
}{J_n(\beta _{jt}R)}=\frac{\beta _{ex}}{\chi _{ex}}\frac{H_n^{(1)^{\prime
}}(\beta _{ex}R)}{H_n^{(1)}(\beta _{ex}R)} 
\end{equation}
describing the normal modes of a cylindrical jet. In equation (A6) the
primes denote derivatives of the Bessel and Hankel functions with
respect to their arguments. The $n$ = 0, 1, 2, 3, 4, etc. modes involve
pinching, helical, elliptical, triangular, rectangular, etc.
distortions of the jet, respectively.

Inside the jet, the azimuthal and axial fluid displacements are related to
the pressure perturbation by
\begin{equation}
\label{A7}\xi _\phi ^{jt}(r)=\pm \frac i{\chi _{jt}}\frac nrp_{1jt}^{*}(r)%
\text{ , and }\xi _x^{jt}(r)=\frac i{\chi _{jt}}k\left[ \frac{\chi _{jt}}{%
\chi _{jt}+(B_{jt}^2/4\pi )(\omega -ku)^2/a_{jt}^2}\right] p_{1jt}^{*}(r)%
\text{ .} 
\end{equation}
With the radial dependence of the total pressure perturbation written
in the form
\begin{equation}
\label{A8}p_{1jt}^{*}(r)=\frac{J_n(\beta _{jt}r)}{J_n(\beta _{jt}R)}%
p_{1jt}^{*}(R) 
\end{equation}
the radial dependence of the radial, azimuthal, and axial displacement
perturbations can all be written in terms of the radial displacement
perturbation appropriate to wave mode $n$ at the jet surface:
\begin{equation}
\label{A9}
\begin{array}{c}
\xi _{r,n}^{jt}(r)= 
\frac{J_n^{^{\prime }}(\beta _{jt}r)}{J_n^{^{\prime }}(\beta _{jt}R)}\xi
_{r,n}^{jt}(R)\text{ ,} \\ \xi _{\phi ,n}^{jt}(r)=\pm n\frac i{\beta _{jt}r} 
\frac{J_n(\beta _{jt}r)}{J_n^{^{\prime }}(\beta _{jt}R)}\xi _{r,n}^{jt}(R)%
\text{ ,} \\ \xi _{x,n}^{jt}(r)=kr\left[ \frac{\chi _{jt}}{\chi
_{jt}+(B_{jt}^2/4\pi )(\omega -ku)^2/a_{jt}^2}\right] \frac i{\beta _{jt}r}%
\frac{J_n(\beta _{jt}r)}{J_n^{^{\prime }}(\beta _{jt}R)}\xi _{r,n}^{jt}(R)%
\text{ ,} 
\end{array}
\end{equation}
where we have used equation (A5) to write
$$
\xi _{r,n}^{jt}(r)= \frac{J_n^{^{\prime
}}(\beta _{jt}r)}{J_n^{^{\prime }}(\beta _{jt}R)}\xi _{r,n}^{jt}(R)
=\frac{\beta _{jt}}{\chi _{jt}}\frac{J_n^{^{\prime
}}(\beta _{jt}r)}{J_n(\beta _{jt}R)}p_{1jt}^{*}(R)\text{ .} 
$$
Displacements of jet fluid from an initial position $(r,\phi ,x)$ are
given by {\boldmath $\xi$}$(r,\phi ,x)=$ {\boldmath $\xi$}$(r)\exp
[i(kx\pm n\phi -\omega t)]$ with the {\boldmath $\xi$}$(r)$ given by
equations (A9), and $\beta _{jt}=\beta _{jt}(\omega ,k)$ and $\chi
_{jt}=\chi _{jt}(\omega ,k)$ where $\omega $ and $k$ are normal mode
solutions of the dispersion relation (A6). The velocity components
associated with these displacements are given by {\bf u}$ _1(r,\phi
,x)=d${\boldmath$ \xi$}$/dt=-i(\omega-ku)${\boldmath
$\xi$}$(r,\phi ,x)$. In general {\boldmath $\xi$}$(r)$ is complex and
{\boldmath $\xi$}$(r,\phi ,x)$ can be written in the form {\boldmath
$\xi$}$(r,\phi _s,x_s)={\bf A}(r)e^{i{\bf \Delta } (r)}\xi
_{r,n}^{jt}(R)\exp [i(kx_s\pm n\phi _s-\omega t)]$ where $\phi _s$ and
$x_s$ are now the azimuthal angle and axial position at the jet
surface, and
\begin{equation}
\label{A10}
\begin{array}{c}
A_r(r)e^{i\Delta _r(r)}\equiv 
\frac{J_n^{^{\prime }}(\beta _{jt}r)}{J_n^{^{\prime }}(\beta _{jt}R)} \\ 
A_\phi (r)e^{i(\Delta _\phi (r)\pm \pi /2)}\equiv n\frac 1{\beta _{jt}r} 
\frac{J_n(\beta _{jt}r)}{J_n^{^{\prime }}(\beta _{jt}R)}e^{\pm i\pi /2} \\ 
A_x(r)e^{i(\Delta _x(r)+\pi /2)}\equiv kr\left[ \frac{\chi _{jt}}{\chi
_{jt}+(B_{jt}^2/4\pi )(\omega -ku)^2/a_{jt}^2}\right] \frac 1{\beta _{jt}r}%
\frac{J_n(\beta _{jt}r)}{J_n^{^{\prime }}(\beta _{jt}R)}e^{i\pi /2} 
\end{array}
\end{equation}
where we have set $\pm i=e^{\pm i\pi /2}$. Thus, we see that
displacements are modified in amplitude by a factor $A(r)$ relative to
those at the jet surface and rotated in azimuthal angle or shifted
along the jet axis by $\Delta (r)$.

To make a transformation to Cartesian coordinates such as are used in
the numerical simulations we will set $\exp \left( i\theta \right)
=\cos \theta $ with $\cos \theta =1$ along the $z$-axis where the flow is
along the $x$-axis.  In this case a point at ($x_0=x_s$, $y_0=-r\sin \phi
_s$, $z_0=r\cos \phi _s$ ) is displaced ($\delta $x, $\delta $y, $\delta
$z) by
\begin{equation}
\eqnum{A11a}\delta x=A_x(r)\xi _{r,n}^{jt}(R)\cos \left[ kx_s-\omega t\pm
n\phi _s+\Delta _x(r)+\pi /2\right] \text{ ,} 
\end{equation}
\begin{equation}
\eqnum{A11b}
\begin{array}{c}
\delta y=-A_r(r)\xi _{r,n}^{jt}(R)\cos \left[ kx_s-\omega t\pm n\phi
_s+\Delta _r(r)\right] \sin \phi _s \\ 
\text{~~~~~}-A_\phi (r)\xi _{r,n}^{jt}(R)\cos \left[ kx_s-\omega t\pm n\phi
_s+\Delta _\phi (r)\pm \pi /2\right] \cos \phi _s\text{ ,} 
\end{array}
\end{equation}
\begin{equation}
\eqnum{A11c}
\begin{array}{c}
\delta z=A_r(r)\xi _{r,n}^{jt}(R)\cos \left[ kx_s-\omega t\pm n\phi
_s+\Delta _r(r)\right] \cos \phi _s \\ 
\text{~~~~~}-A_\phi (r)\xi _{r,n}^{jt}(R)\cos \left[ kx_s-\omega t\pm n\phi
_s+\Delta _\phi (r)\pm \pi /2\right] \cos \phi _s\text{ .} 
\end{array}
\end{equation}
\setcounter{equation}{11}
Of particular interest are distortions to the initially circular jet
cross section. In the numerical simulation a cross section is obtained
at a fixed axial distance and time, say at $x_s=0$ and $t=0$, and the
cross section of a displacement surface can therefore be constructed by
obtaining $\delta x(r,\phi _s,x_s)=\delta x(r,\phi _s,0)$ from equation
(A11a) and then by obtaining $y_0+\delta y$ and $z_0+\delta z$ with
$x_s\rightarrow -\delta x(r,\phi _s)$ from equations (A11b \& A11c).
This axial shift invokes a small additional azimuthal angle or axial
shift of $-k_R\delta x(r,\phi _s)$ in addition to that arising from
$\Delta (r)$. If displacements associated with many different surface
and body wave modes are assumed to add in a linear fashion but each
mode is oriented differently in the $y-z$ plane, then we may set $\phi
_s=\phi _s+\Delta _{nm}$ in equations (A11) where $\Delta _{nm}$
provides the azimuthal or axial shift relative to the $z$-axis or
$x$-axis, respectively, for the appropriate surface or body wave mode.
In Cartesian coordinates the total velocity will be given by ${\bf
u}(x,y,z)= {\bf u}_0+{\bf u}_1(x,y,z)={\bf u}_0-i(\omega-ku)${\boldmath
$\delta$ }$(r,\phi _s,x_s)$ where ${\bf u}_0$ is the jet velocity along
the $x$-axis, the components of {\boldmath $\delta$}$(r,\phi _s,x_s)$
are given by equations (A11), and $(x,y,z)=(x_s+\delta x$, $y_0+\delta
y$, $z_0+\delta z)$.

\newpage

\figcaption{
Toroidal component and helical pitch of the magnetic field as a function of
the jet radius.
}

\figcaption{
Magnetic pressure cross section images from simulation A  at axial
distances from 6R to 48R in 6R increments from top left to bottom
right. The $y$-axis is in the vertical direction, the $z$-axis
(line-of-sight axis) is in the horizontal direction and the flow
direction ($x$-axis) is into the page. Each cross section image has
dimension $10R \times 10R$ and darker indicates higher values of the
magnetic pressure.  Below the cross section images are two images of
dimension $20R\times 60R$ showing (top) total intensity, and (bottom)
fractional polarization where the brightest regions are nearly 70\%
polarized. Markings at 6R intervals show where cross sections and
transverse velocity profiles are located.
}

\figcaption{
Same as Figure 2 but for simulation B.
}

\figcaption{
Same as Figure 2 but for simulation C at time $\tau_d = 34$.
}

\figcaption{
Same as Figure 2 but for simulation C at time $\tau_d = 36$.
}

\figcaption{
Axial profiles of the density, specific thermal energy, sonic Mach
number, $M_{jt}\equiv u/a_{jt}$, and the three velocity components
scaled relative to the sound speed, $a_{ex}$, in the external medium in
simulation A.
}

\figcaption{
Same as Figure 6 but for simulation B.
}

\figcaption{
Same as Figure 6 but for simulation C ($\tau_d = 36$).
}

\figcaption{
Transverse profiles along the $y$-axis of density (left column),
magnetic pressure (center column), and axial velocity (right column) at
axial distances of 12R, 24R, 36R and 48R from top to bottom rows in
simulation A. The density is scaled relative to the ambient density in
the external medium, the magnetic pressure is in units such that
$B^2/2$ is pressure units, and the velocity is scaled relative to the
sound speed, $a_{ex}$, in the external medium.
}

\figcaption{
Same as Figure 9 but for simulation B.
}

\figcaption{
Same as Figure 9 but for simulation C ($\tau_d = 36$).
}

\figcaption{ 
Numerical solution of the dispersion relation for parameters
appropriate to simulation C. The surface (S) wave and first three body
(B$_1$, B$_2$, B$_3$) wave solutions are shown for the helical,
elliptical, triangular and rectangular modes. The dotted lines give the
real part of the wavenumber, k$_R$, and the dashed lines give the
absolute value of the imaginary part of the wavenumber, $\left|
k_I\right| $.
}

\figcaption{ 
Fluid displacement surfaces accompanying the surface and first three
body waves associated with the helical, elliptical, triangular, and
rectangular modes. Displacement surfaces are computed for parameters
appropriate to the heavy jet simulation C using the solutions to the
dispersion relation shown in Figure 12 at the maximum growth rates.
}

\figcaption{ 
Fits to the surface distortions seen in the cross sections in Figure 4
using the helical, elliptical, triangular, and rectangular surface
modes only.
}

\newpage

\begin{table} 
 \begin{center}
 \caption{Initial Conditions}
 \vspace{0.1cm}
 \begin{tabular}{c c c c c l l l l} \hline \hline
     {\bf Simulation}  &  {\bf $\eta$}  &  {\bf $a_{jt}/a_{ex}$}  &  {\bf $V_{A,jt}/a_{ex}$}  &  {\bf $a_{jt}^{ms}/a_{ex}$} &  {\bf $M_{ex}$}   &  {\bf $M_{jt}$}  &  {\bf $M^{A}_{jt}$}   &  {\bf $M^{ms}_{jt}$} \\ \hline
     A   &  0.25  &  1.47  &  1.50  & 2.10  &  12.50  &  8.50  &  8.33  &  5.95  \\
     B   &  1.00  &  0.74  &  0.75  & 1.05  &   5.14  &  7.00  &  6.86  &  4.90  \\
     C   &  4.00  &  0.37  &  0.38  & 0.52  &   4.04  &  11.0  &  10.78 &  7.70  \\ \hline
  \end{tabular}
  \end{center}
\end{table}

\begin{table} 
 \begin{center}
 \caption{Wavelengths at Maximum Growth in Simulation C}
 \vspace{0.1cm}
 \begin{tabular}{l c c c c} \hline \hline
     {\bf Mode}  &  $\lambda_S^*$/R  &  $\lambda_{B1}^*$/R  &  $\lambda_{B2}^*$/R  &  $\lambda_{B3}^*$/R    \\  \hline
     Pinch       &  ---       &  4.9        &  2.5        & 1.7                                     \\
     Helical     &  12.5      &  3.4        &  2.0        & 1.4                                       \\
     Elliptical  &  9.2       &  2.7        &  1.7        & 1.3                                       \\
     Triangular  &  6.5       &  2.3        &  1.5        & 1.2                                        \\
     Rectangular &  4.8       &  2.0        &  1.3        & 1.0                                        \\  \hline
  \end{tabular}
  \end{center}
\end{table}

\begin{table} 
 \begin{center}
 \caption{Observed Surface Mode Parameters in Simulation C}
 \vspace{0.1cm}
 \begin{tabular}{l c c c c} \hline \hline
  {\bf Mode}  &  $\lambda^{obs}/\lambda_S^*$  &  $v_w^{obs}/v^*_{w,S}$  &  $\xi_{rn}^{obs}$  &  $\xi^{max}_{pred}$    \\  \hline
     Helical     &  1.1   &  $\sim 1.1$    &  $\sim 0.8$   & $\sim 0.75$                                       \\
     Elliptical  &  1.5    &  $\sim 1.1$    &  $\sim 0.6$    & $\sim 0.60$                                       \\
     Triangular  &  1.4    &  $\sim 1.1$    &  $\sim 0.1$   & $\sim 0.40$                                        \\
     Rectangular &  1.2    &  $\sim 1.1$    &  $\sim 0.07$   & $\sim 0.25$                                        \\  \hline
  \end{tabular}
  \end{center}
\end{table}

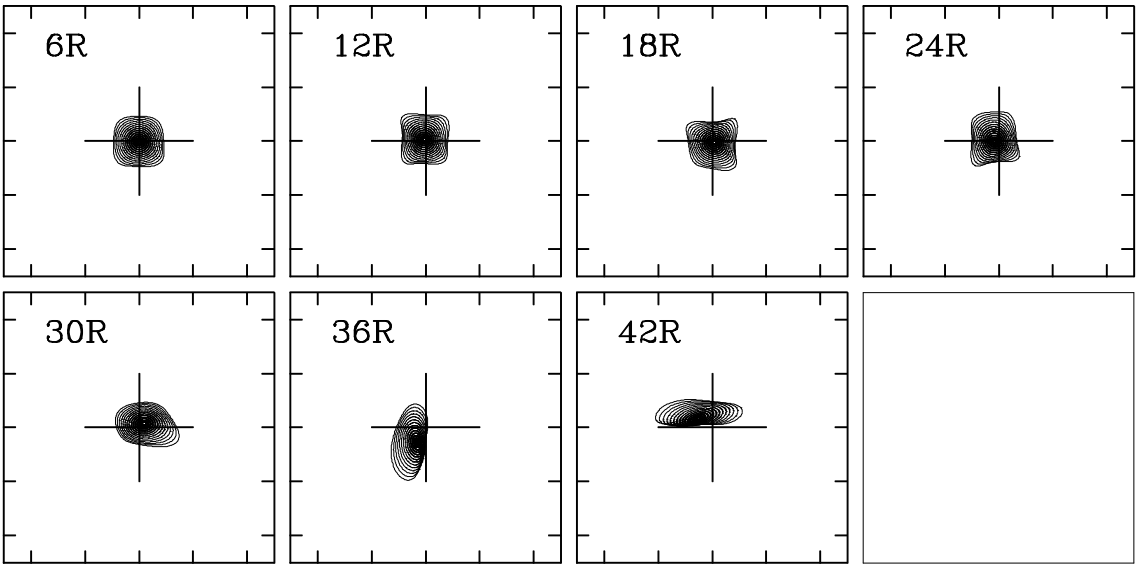
\end{document}